\documentclass[lettersize,journal]{IEEEtran}
\usepackage{amsmath,amsfonts}
\usepackage{algorithmic}
\usepackage{algorithm}
\usepackage{array}
\usepackage[caption=false,font=normalsize,labelfont=sf,textfont=sf]{subfig}
\usepackage{textcomp}
\usepackage{stfloats}
\usepackage{url}
\usepackage{verbatim}
\usepackage{graphicx}
\usepackage{cite}
\usepackage{multirow}
\usepackage{amsfonts}
\usepackage{threeparttable}
\usepackage{color}
\usepackage{makecell}
\usepackage{bm}
\usepackage{epstopdf}
\usepackage{url}
\usepackage{booktabs}

\begin{document}

\title{An Interpretable MRI Reconstruction Network with Two-grid-cycle Correction and Geometric Prior Distillation}

\author{Xiaohong Fan, Yin Yang, Ke Chen, Jianping Zhang, and Ke Dong
\thanks{This work was supported by the National Natural Science Foundation of China (NSFC) under Grants 11771369, 12071402, 12261131501, and the National Key Research and Development Program of China under Grant 2020YFA0713503, also partly by grants from {the Education Bureau of Hunan
Province in China under Grant 22A0119}, and the Project of Scientfic Research Fund of the Hunan Provincial Science and Technology Department under Grants 2020JJ2027, 2022RC3022, 2020ZYT003, and Postgraduate Scientific Research Innovation Project of Hunan Province (Grant no. CX20210598) and Postgraduate Scientific Research Innovation Project of  Xiangtan University (Grant no. XDCX2021B097), P. R. China. (Corresponding author: Jianping Zhang).}
\thanks{X. Fan is with the School of Mathematics and Computational Science, Xiangtan University, and Key Laboratory of Intelligent Computing \& Information Processing of Ministry of Education, Xiangtan 411105, China (fanxiaohong@smail.xtu.edu.cn).}
\thanks{Y. Yang is with the School of Mathematics and Computational Science, Xiangtan University, and Hunan National Applied Mathematics Center, and Hunan Key Laboratory for Computation and Simulation in Science and Engineering, Xiangtan 411105, China ( yangyinxtu@xtu.edu.cn).}
\thanks{K. Chen is with
Centre for Mathematical Imaging Techniques and Department of Mathematical Sciences, The University of Liverpool, Liverpool, Merseyside L6972L, United Kingdom (k.chen@liverpool.ac.uk).}
\thanks{J. Zhang is with the School of Mathematics and Computational Science, Xiangtan University, and Hunan National Applied Mathematics Center, and Key Laboratory of Intelligent Computing \& Information Processing of Ministry of Education, Xiangtan 411105, China. (jpzhang@xtu.edu.cn).}
\thanks{K. Dong is with Department of Radiation Oncology,
Xiangtan Central Hospital, Xiangtan 411101, China (e-mail: kedong116@yahoo.com).}
\thanks{-- Accepted to Biomedical Signal Processing and Control, March, 2023.}
}

\markboth{
}%
{Shell \MakeLowercase{\textit{et al.}}: A Sample Article Using IEEEtran.cls for IEEE Journals}

\IEEEpubid{
}

\maketitle

\begin{abstract}
Although existing deep learning compressed-sensing-based Magnetic Resonance Imaging (CS-MRI) methods have achieved considerably impressive performance, explainability and generalizability continue to be challenging for such methods since the transition from mathematical analysis to network design not always natural enough, often most of them are not flexible enough to handle multi-sampling-ratio reconstruction assignments. {In this work, to tackle explainability and generalizability, we propose a unifying deep unfolding multi-sampling-ratio interpretable CS-MRI framework.} The combined approach offers more generalizability than previous works whereas deep learning gains explainability through a geometric prior module. Inspired by the multigrid algorithm, we first embed the CS-MRI-based optimization algorithm into correction-distillation scheme that consists of three ingredients: pre-relaxation module, correction module and geometric prior distillation module. Furthermore, we employ a condition module to learn adaptively step-length and noise level, which enables the proposed framework to jointly train multi-ratio tasks through a single model. { The proposed model not only compensates for the lost contextual information of reconstructed image which is refined from low frequency error in geometric characteristic k-space}, but also integrates the theoretical guarantee of model-based methods and the superior reconstruction performances of deep learning-based methods. Therefore, it can give us a novel perspective to design biomedical imaging networks. { Numerical experiments show that our framework outperforms state-of-the-art methods in terms of qualitative and quantitative evaluations.} {Our method achieves 3.18 dB improvement at low CS ratio 10\% and average 1.42 dB improvement over other comparison methods on brain dataset using Cartesian sampling mask.} 

\end{abstract}

\begin{IEEEkeywords}
Deep learning, CS-MRI reconstruction, Unfolding explainable network, Two-grid-cycle correction, Geometric prior distillation, Multi-sampling-ratio reconstruction.
\end{IEEEkeywords}

\section{Introduction}
Magnetic Resonance Imaging (MRI) is one kind of widely used medical imaging modalities for clinical diagnosis, which requires a long scan time and has the risk of motion-related artifacts in the reconstructed image.
CS-MRI that only requires much lower sampling rate than Nyquist sampling theory has been proposed to reconstruct image from the sparse characteristics of signals. Although it may cause aliasing artifacts in a spatial domain, the quality of reconstructed image is not significantly reduced.

Classical CS-MRI methods can learn directly more flexible sparse representation from under-sampled data by restricting solution in a specific transformation domain \cite{Song2016,KYCui2021a} or in a generic dictionary-based subspace \cite{Zhan2016}. Discrete wavelet transform \cite{Lai2016} and discrete cosine transform \cite{Lingala2013} have been used for CS-MRI reconstruction. Due to simple and effective, Total Variation (TV) regularization has been widely used in MRI reconstruction although it introduces staircase artifacts in reconstructed image \cite{Block2007}. Some sophisticated non-local sparsity methods that use groups of local similar patches to exploit the non-local self-similarity properties, can capture more texture priors and improve significantly CS reconstruction performance, e.g., patch-based directional wavelets (PBDW) \cite{Qu2012}, patch-based nonlocal operator (PANO) \cite{Qu2014}. The BM3D denoiser \cite{Metzler2016} has been integrated into CS reconstruction as a new approximate message passing (AMP) framework. Many studies formulate CS a reconstruction problem as the sparsity-regularized optimization problem and then solve it by using different iterative algorithms such as Iterative Shrinkage Threshold Algorithm (ISTA) \cite{Beck2009}, Primal-Dual algorithm \cite{Chambolle2010} and Alternating Direction Multiplier Method (ADMM) \cite{Boyd2010} etc. All the above methods that are based on strongly interpretable and well predefined sparsity image prior, have the advantages of theoretical support and strong convergence. {However, they usually require expensive computations and face the problems of selecting suitable regularizers and model parameters. Consequently, the reconstructed results of the classical CS-MRI methods are usually unsatisfactory.}

Recently, deep learning (DL) has achieved great success in computer vision community \cite{Datta2022}. A deep Convolutional Neural Network (CNN) has been proposed to reconstruct MRI from under-sampled data \cite{Wang2016b}.
Based on learning a direct inversion of image inverse problem, the network \cite{Jin2017} combines multi-resolution decomposition and residual learning to remove artifacts while preserving image structure. RefineGAN is a variant of fully-residual convolutional auto-encoder and generative adversarial networks (GANs) with cyclic data consistency loss for CS-MRI \cite{Quan2018}. DAGAN couples adversarial loss with an innovative content loss to better preserve texture and edges in CS-MRI, and also incorporates frequency-domain information to enforce similarity in both the image and frequency domains \cite{Yang2018}. A generalized double-domain GAN is exploited to reconstruct undersampled multi-contrast MRI \cite{HNWei2022a}. {SynDiff based on adversarial diffusion modeling is proposed to improve performance in medical image translation \cite{Oezbey2022}. The first adaptive diffusion prior for MRI reconstruction named AdaDiff is proposed to improve performance and reliability against domain shifts \cite{Guengoer2022}.} {W-net can work both in $k$-space and in image domain for CS-MRI} \cite{Souza2019}, which is composed of a complex-valued residual U-net in $k$-space, an iFFT operation, and a real-valued U-net in image domain. DuDoRNet with deep embedded T1 prior simultaneously recovers $k$-space information and image for accelerating the acquisition of MRI \cite{Zhou2020}. These existing DL-based CS-MRI methods are data-driven based on a large amount of training data  and without any model prior.

Deep unfolding learnable framework of inheriting the merits of model-based and DL-based CS-MRI methods, has sufficient theoretical support and also good performance \cite{Yang2020,Wang2021,Aghabiglou2022}. It is first proposed to learn optimal sparse codes in the Learned Iterative Shrinkage-Thresholding Algorithm (LISTA) \cite{Gregor2010}. Later, Yang et al. \cite{Yang2016} presented a novel deep ADMM-Net framework to supervise data flow graph in image reconstruction network. A cascaded dilated dense network with two-step data consistency operation in $k$-space is designed for CS-MRI reconstruction \cite{Zheng2019}. {A novel parallel-stream fusion model (PSFNet) synergistically fuses scan-specific and scan-general priors for performant MRI reconstruction from Scratch in low-data regimes \cite{Dar2023}.} Also a conjugate gradient image reconstruction with a CNN-based regularization prior is employed to build MoDL architecture \cite{Aggarwal2019}. Then by unfolding the iterative process of variable splitting optimization scheme, Duan et al. proposed a novel end-to-end trainable deep neural network (DNN) denoted as VS-Net \cite{Duan2019}. DC-CNN \cite{Schlemper2018} using a deep cascade of CNNs with data consistency layers is proposed to reconstruct MRI from under-sampled data. ISTA-Net+ is designed by mapping ISTA into deep CNN framework to learn proximal mapping \cite{Zhang2018}. To embedded FISTA algorithm \cite{Lustig2007} into a deep network, FISTA-Net \cite{Xiang2020} that consists of three-step update blocks including gradient descent, proximal mapping, and acceleration is designed.

Recently, the techniques of solving linear/nonlinear system have been widely used to design effective DNNs. ResNet \cite{He2016} is partly motivated by the hierarchical correction of residual in classical iterative algorithm \cite{He2016a}. PolyNet designs a PolyInception module to enhance feature extraction of network \cite{Zhang2017}. RevNet\cite{Gomez2017} and LM-ResNet \cite{Lu2018} can be interpreted as different reversible Euler-type discrete dynamic systems of ordinary differential equations (ODEs). A deep multigrid method is proposed to optimize restriction and prolongation operations in two-grid scheme, and is straightforwardly extended to the geometric multigrid method \cite{Katrutsa2017}. A multigrid extension of CNNs is proposed to improve accuracy and computational efficiency on CIFAR and ImageNet classification tasks \cite{Ke2017}. MgNet is also designed for image classification to explore the connection between multigrid and CNNs \cite{He2019}. These models can be effective for improvements of deep learning models, and in particular for the mathematical understanding and analysis of network architecture.

However, there are still two main shortcomings: 1) the derivation from mathematical theory to network design is not always natural enough for these existing deep unfolding methods. ISTA-Net and FISTA-Net directly replace the nonlinear transformation by several convolution layers, but no reasonable explanation is given. The proximal-point sub-problem has not well been analyzed and explained in DC-CNN, MoDL, and VS-Net. It leads that CNN is directly used to reconstruct the low CS-ratio image;
2) most of existing CS-MRI methods are not flexible enough to handle multi-sampling-ratio reconstruction assignments. For each CS ratio, these methods usually have to train an independent model by considering it as a single task, which results in expensive training cost and large storage space. And this is often contradicted with the fact that real scenarios often contain different CS ratios.

To overcome such drawbacks, inspired by efficient correction of multigrid technique in multiple resolutions, we start from classical CS-MRI optimization problem, and embed it into a two-scale correction-distillation architecture which consists of pre-relaxation, correction and geometric prior distillation. { 
Firstly, the derivation from mathematical theory to network design is more natural than the existing deep unfolding methods, to deliver the expected generalizability of deep learning.
Secondly,
we learn the proximal-point sub-problem to distill features of different geometric characteristic domains.   Finally, the proposed framework can handle multi-sampling-ratio CS-MRI tasks in a single model to avoid expensive training cost and large storage, while the existing CS-MRI methods have to train an independent model for each CS ratio.}
 The major differences between this work and the previous work \cite{Fan2021} are that we propose an interpretable DL framework to train/predict multi-sampling-ratio CS-MRI tasks in a single model, and two-grid-cycle correction and geometric priors are refined to design network architecture for further promoting reconstruction performance. The main contributions of this work can be summarized as follows
\begin{itemize}
        \item[1)] A novel deep unfolding unified framework which enjoys much flexibility to conduct multi-sampling-ratio CS-MRI through a single model is proposed.
        \item[2)] A multi-grid inspired unfolding correction-distillation scheme which can not only incorporate frequency-domain information to compensate for low-frequency error in $k$-space, but also learn geometric priors of MR image by adding a geometric prior distillation module. The proposed method gives us a novel perspective on designing biomedical imaging network architectures.
        \item[3)] A condition module is employed to transmit CS sampling ratio to step-length and to characterize noise level in every stage.
        \item[4)] A linkage of model-based and learning-based methods which integrates the theoretical guarantee of model-based methods and the superior reconstruction performances of DL-based methods, to deliver the expected generalizability of deep learning.
        \item[5)] A learning framework to make all physical-model parameters learnable to ensure that suitable choices are automated, as shown in extensive experiments in favourably comparing with state-of-the-art methods in terms of visualizations and quantitative evaluations on flexibility and stability.
\end{itemize}

The rest of paper is organized as follows. In Section II,
the proposed correction-distillation CS-MRI framework is introduced in detail. The
experimental results are shown in Section III. The conclusion of
this paper is presented in Section IV.

\begin{figure*}
        \centerline{\includegraphics[width=2\columnwidth]{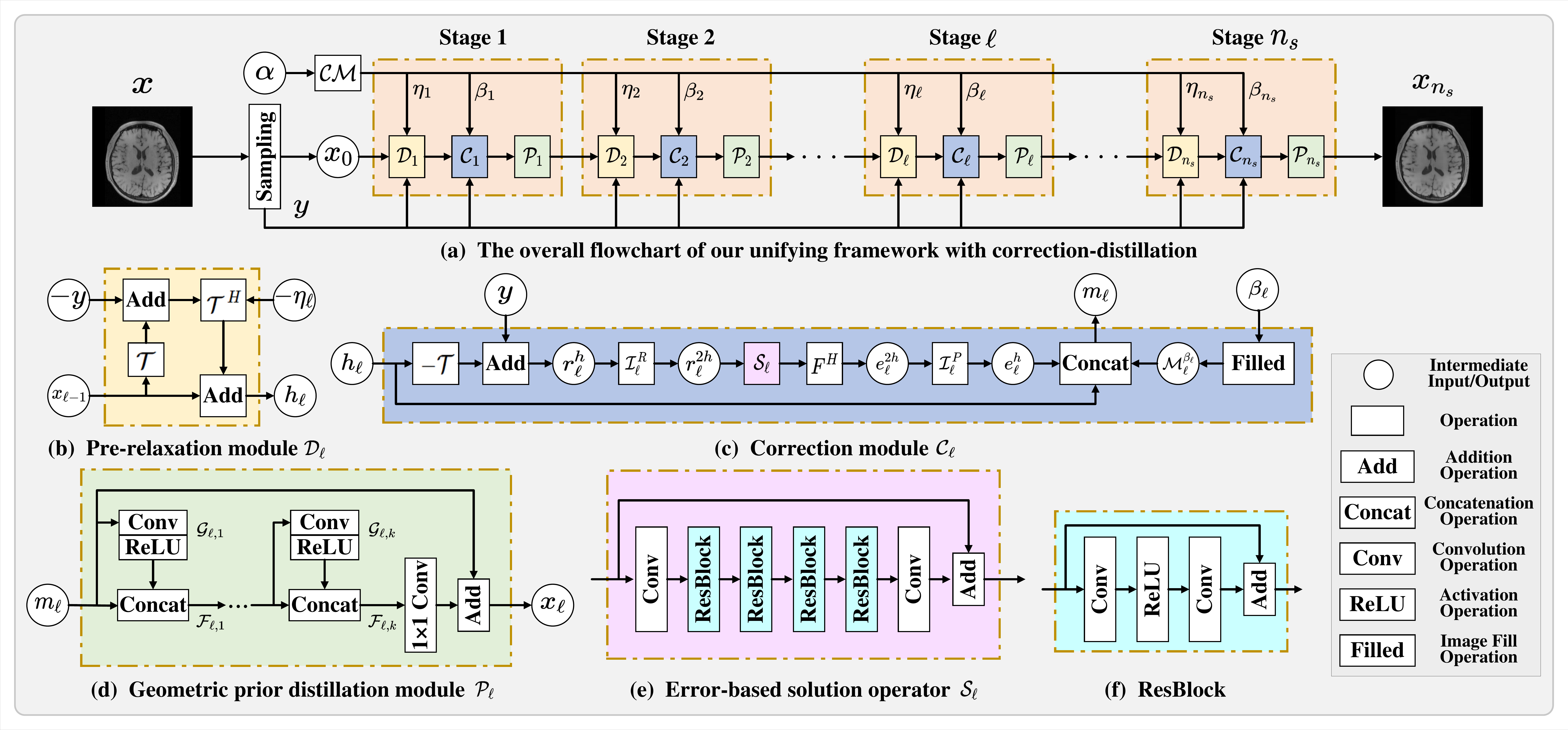}}
        \caption{The overall architecture of the proposed unifying multi-sampling-ratio CS-MRI framework with two-grid-cycle correction and geometric prior distillation (CGPD-CSNet). It consists of three major ingredients, i.e. pre-relaxation module $\mathcal{D}_\ell$, correction module $\mathcal{C}_\ell$, and geometric prior distillation module $\mathcal{P}_\ell$.}
        \label{fig_architecture}
\end{figure*}

\section{Methodology}

The general sampling equation in Fourier space can be formulated as follows
\begin{equation}
        \mathcal{T}\bm{x}=\bm{y},
        \label{eq0}
\end{equation}
where $\bm{x} \in \mathbb{R}^{d}$ ($d=mn$) is a target MRI, $\mathcal{T}=PF\in \mathbb{R}^{q\times d} $ is an under-sampled Fourier measurement matrix, $P$ is an under-sampling matrix and $F$ is a discrete Fourier transform, $\bm{y} \in \mathbb{R}^{q} (q\ll d)$ is the under-sampled $k$-space data. It is well known that the problem (\ref{eq0}) is ill-posed,
%
therefore, the classical CS-MRI optimization of image reconstruction can be given by
\begin{equation}
        \min_{\bm{x}}\left\{\mathcal{E}(\bm{x}):= \mathcal{H}(\bm{x};\bm{y})+\gamma \mathcal{R}(\bm{x})\right\},
        \label{eq1}
\end{equation}
where $\mathcal{H}(\bm{x};\bm{y})=\frac{1}{2}\left\|\mathcal{T} \bm{x}-\bm{y}\right\|_{2}^{2}$ is a data fidelity term, $\mathcal{R}(\bm{x})$ is a regularizer with either a known or unknown geometric prior, $\gamma$ is a regularization parameter that balances data fidelity term and regularizer.

In this section, we aim to design an explainable unifying deep learning CS-MRI reconstruction framework, which is denoted as CGPD-CSNet. Our method inherits the main advantages of model-based and DL-based methods. The overall architecture of the proposed network, which is inspired for learning a unified multi-sampling-ratio CS-MRI minimization, is shown in Fig.\ref{fig_architecture}, and more details are provided hereafter.

\subsection{Proposed Learnable CS-MRI Unrolled framework}
The problem presented in (\ref{eq1}) can be effectively resolved using a two-step optimization procedure, which not only avoids time-consuming numerical optimization  but also reconstructs more spatial-frequency geometric priors. In details, we split (\ref{eq1}) into two subproblems as follows
\begin{description}
        \item[(P1)] linear reconstruction subproblem:
        \[\bm{m}_{\ell}\in \{\bm{x}: \mathcal{T} \bm{x}=\bm{y}\};\]
        \item[(P2)] geometric prior reconstruction subproblem:
        \[\bm{x}_{\ell}:=\underset{\bm{x}}{\arg \min } \left\{\frac{1}{2}\|\bm{x}-\bm{m}_{\ell}\|_{2}^{2}+\lambda_\ell\mathcal{R}(\bm{x})\right\}.\]
\end{description}

To learn more detailed information about many specific features that are invariant to contextual information and uninformative
intensity-variations, we consider a generative feature correction
module to solve the linear system (\ref{eq0}),
which involves a two-grid-cycle scheme of classical multi-grid algorithm \cite{McCormick1987,KChen05} with an initial {value} $\bm{\bar{x}}^{h}$ and right-hand side $\bm{y}^{h}$ on fine grid $h$ denoted by
$$
\bm{x}^{h} \leftarrow \text{MG}^{h}\left(\bm{\bar{x}}^{h}, \bm{y}^{h},\mathcal{T}^h\right),
$$
and the above linear reconstruction solution can be given by the following steps:\\
\textbf{Step 1. (pre-relaxation)}: Perform $\nu_0$ sweeps to approximate the solution $\bm{x}^h_l$ of $\mathcal{T}^{h} \bm{x}^{h}=\bm{y}^{h}$ with initial {value} $\bm{\bar{x}}^{h}$;\\
\textbf{Step 2. (coarse grid solution)}: Compute $\bm{r}^{h}=\bm{y}^{h}-\mathcal{T}^{h} \bm{x}^{h}_l$ and $\bm{r}^{2h}=I_{h}^{2h} \bm{r}^{h}$, and then seek an approximation $e^{2 h}$ to the solution of $\mathcal{T}^{2h}\bm{e}^{2h}=\bm{r}^{2h}$;\\
\textbf{Step 3. (coarse grid correction)}: Let $\bm{x}^{h}_m \leftarrow \bm{x}^{h}_l+I_{2h}^{h}\bm{e}^{2h}$;\\
\textbf{Step 4. (post-relaxation)}: Run $\nu_1$ sweeps to approximate the solution $\bm{x}^h_r$ of $\mathcal{T}^{h}\bm{x}^{h}=\bm{y}^{h}$ with initial {value} $\bm{x}^{h}_m$, \\
where $\nu_0$ and $\nu_1$ are the numbers of relaxation sweeps to be done before and after nested process on fine grid $\Omega^{h}$, respectively. { The update solutions with subscripts $l$ and $m$ denote the results of the pre-relaxation step and coarse grid correction step, respectively. $h$ and $2h$ represent the fine and coarse grid cycles, respectively.} $I_{h}^{2h}$ is a restriction operator which restricts residual from fine to coarse grid, and $I_{2h}^{h}$ is a prolongation operator which restricts the corrected residual from coarse to fine grid.

In the following we  bridge the correspondences between well established
principles in conventional modeling methods and CNN reconstruction
networks.

\subsubsection{Proposed correction-distillation architecture}
Inspired by the excellent performance of multigrid in multi-scale error correction, we propose a learnable correction-distillation framework which is embedded to quickly extract low frequency increment in $k$-space (\emph{see} Fig.\ref{fig_architecture}(a)). At each stage $\ell$, we thus construct two modules based on \textbf{Steps 1 -- 3} to generate
solution iterations for  subproblem (P1) as follows
\begin{align}
        \bm{h}_{\ell}&=\bm{x}_{\ell-1}-\eta_{\ell}{\mathcal{T}^{H}}(\mathcal{T}\bm{x}_{\ell-1}-\bm{y}),
        \label{eq4}\\
        \bm{e}^{2h}_\ell&={{F}^{H}}\mathcal{S}_{\ell}\mathcal{I}_{\ell}^{R}(\bm{y}-\mathcal{T} \bm{h}_{\ell}),
        \label{eq5}\\
        \bm{m}_{\ell}&=\text{Concat}(\bm{h}_{\ell},\mathcal{I}_{\ell}^{P}\bm{e}^{2h}_\ell,\mathcal{M}_\ell^{\beta_\ell}),
        \label{eq6}
\end{align}
where { $\bm{x}_0$ is computed by the transformation $\bm{x}_0=\mathcal{T}^{H}\bm{y}$ as one of the initial inputs of the Stage 1 and $\nu_0=1$}, $\eta_{\ell}$ is a learnable step-length, { ${F}^{H}$ is an inverse discrete Fourier transform.} $\text{Concat}(\cdot,\cdot,\cdot)$ is a correction operation, and $\mathcal{M}_\ell^{\beta_\ell}$ is filled with noise level $\beta_\ell$ learned from condition module $\mathcal{CM}$. 

{ To solve $\mathcal{T}^{h} \bm{x}^{h}=\bm{y}^{h}$ or subproblem (P1) at Stage $\ell$, we only denote the inner iteration update in two-grid-cycle scheme \textbf{Steps 1 -- 4}, where the initial value $\bar{\bm{x}}_m$ in the pre-relaxation step of the Stage $\ell$ (outer iteration) is the output $\bm{x}_{\ell-1}$ of the previous Stage $\ell-1$, and the initial value in the post-relaxation step of Stage $\ell$ (outer iteration) is the update $\bm{x}^h_m$ of the coarse grid correction step. In fact, these initial values of other Stages don't need to be manually given except Stage 1.}

More importantly, we observed that correction step (\ref{eq5}) with error-based solution operator $\mathcal{S}_\ell$ (\emph{see} Fig.\ref{fig_architecture}(e)) can effectively learn significant details to yield improved solution of subproblem (P1). It turns out that the results have almost no significant differences with or without using the post-relaxation \textbf{Step 4}. To this end, we set the unifying CS-MRI framework without post-relaxation as default in this work. 

However, if we incorporate the learnable geometric prior distillation stage (P2) into the solution fidelity of image inverse problem (\ref{eq0}), 
the network can learn the expected MR image priors. For this purpose, we add the geometric prior fidelity module (\emph{see} Fig.\ref{fig_architecture}(d)) in our network, which is designed by solving the equation
\begin{align}
(I-\mathcal{N})(\bm{x}_{\ell})=:\bm{x}_{\ell}-\mathcal{N}(\bm{x}_{\ell})=\bm{m}_{\ell},
\label{eq7}
\end{align}
where the nonlinear function $\mathcal{N}(\bm{x})=-\lambda\frac{\partial\mathcal{R}(\bm{x})}{\partial\bm{x}}$ denotes the geometric characteristics of $\bm{x}$.

The above correction-distillation architecture (\ref{eq4}) - (\ref{eq7}) can be performed by a pre-relaxation module $\mathcal{D}_\ell$, a correction module $\mathcal{C}_\ell$, and a geometric prior distillation module $\mathcal{P}_\ell$. It's well known that the module $\mathcal{D}_\ell$ often results in heavy artifacts, while the correction module $\mathcal{C}_\ell$ is used to quickly generate the low frequency correction in $k$-space. Then the explainable module $\mathcal{P}_\ell$ is learned to produce more texture priors.

\subsubsection{Pre-relaxation module $\mathcal{D}_\ell$} At each stage $\ell$,
to provide the basis-solution guarantee in high-frequency layer, we can obtain an approximate solution $\bm{h}_{\ell}$ of subproblem (P1)
by solving the least-square problem
\begin{equation*}\label{least-square_prob}
        \min\limits_{\bm{x}} \left\{\mathcal{H}(\bm{x};\bm{y})=\frac{1}{2}\left\|\mathcal{T} \bm{x}-\bm{y}\right\|_{2}^{2}\right\}.
\end{equation*}
Especially, the gradient descent \eqref{eq4} with the parameter $\eta_\ell$ is implemented to perform
a linear reconstruction, which is also called as the pre-relaxation module $\mathcal{D}_\ell$ defined by
\begin{align}
        \begin{split}
                \bm{h}_{\ell}&=\mathcal{D}_\ell(\bm{x}_{\ell-1},\eta_{\ell},\bm{y},\mathcal{T})\\
                &=\bm{x}_{\ell-1}-\eta_{\ell}{ \mathcal{T}^{H}}(\mathcal{T} \bm{x}_{\ell-1}-\bm{y}). 
        \end{split}
        \label{eq8}
\end{align}

The module $\mathcal{D}_\ell$ corresponding to Eq.(\ref{eq8}) is directly used to generate the preliminary approximation from the previous $\bm{x}_{\ell-1}$ (\emph{see} Fig.\ref{fig_architecture}(b)). { Setting $\bm{x}_0=\mathcal{T}^{H}\bm{y}$ as the initial guess of Stage 1 (not all stages) has been widely used in MRI reconstruction and CS tasks \cite{Qu2012,Qu2014,Xiang2020}. In this work, we also suggest that $\bm{x}_0=\mathcal{T}^{H}\bm{y}$ is an input of the Stage 1 and $\bm{x}_{\ell-1}$ is one of the inputs of Stage $\ell$}. It is well known that the step-length $\eta_\ell$ should be positive, and decrease smoothly as the increase of iterations in traditional model-based methods. { While the classical model-based methods usually face the problems of either hardly selecting suitable fixed step-length $\eta_\ell$ or requiring expensive computation costs determined by line search. Consequently, the reconstructed results of the classical CS-MRI methods are usually unsatisfactory. Setting the step-length $\eta_\ell$ to be learnable, which has been widely used in deep unfolding methods \cite{Zhang2018,Xiang2020,You2021}, is reasonable and appropriate. Especially, the parameters of the learnable step-length $\eta_\ell$ (429 parameters) is negligible in comparing with the total parameters (4.4 million parameters) in the proposed framework. Moreover, such a learned model could implement multiple CS ratios CS-MRI throungh a single model when processing medical data with multi-sampling-ratio.} To enhance network flexibility, we set the step-length $\eta_\ell$ to be
learnable during iterations. There are a variety of ways to use training data to learn step-length $\eta_\ell$. Here, we adopt the learned step-length $\eta_{\ell}$ from condition module $\mathcal{CM}$ in \S \ref{section_CM}.

\subsubsection{Correction module $\mathcal{C}_\ell$} The proposed correction module corresponds strictly to a composition of coarse grid solution \eqref{eq5} and coarse grid correction \eqref{eq6} in two-grid-cycle update scheme. The implementation process of solution correction $\mathcal{C}_\ell$ is shown in Fig.\ref{fig_architecture}(c).

Similar to {\bf Step 2} in classical two-grid-cycle scheme, we first design the reconstruction residual block in $k$-space as
\[\bm{r}_\ell^{h}=\bm{y}^h-\mathcal{T}^h\bm{h}^h_{\ell}.
\]
Then we restrict the reconstructed $k$-space residual $\bm{r}_\ell^{h}$ on fine level to $\bm{r}_\ell^{2h}$ on the coarse level, which is denoted by
\[\bm{r}_\ell^{2h}=\mathcal{I}_\ell^{R}\bm{r}_{\ell}^{h}=\mathcal{I}_\ell^{R}(\bm{y}^h-\mathcal{T}^h\bm{h}^h_{\ell}):=\mathcal{I}_\ell^{R}(\bm{y}-\mathcal{T}\bm{h}_{\ell}),\]
where a convolution $\mathcal{I}_\ell^{R}$ with stride 2 pixel { and $2\times2$ kernel} is learnt to exploit the restriction operator $I_{h}^{2 h}$. Note that $\mathcal{I}_\ell^{R}$ and $I_{h}^{2 h}$ have the similar meaning but the former is for learning implementation while the latter is for mathematical description.

To compute a more accurate correction $\bm{e}^{2 h}_\ell$ as the solution of $\mathcal{T}^{2h}_\ell \mathbf{e}^{2h}_\ell=\bm{r}^{2h}_\ell$ and reduce
the computational costs, we propose
to learn an error-based solution operator $\mathcal{S}_{\ell}$ to quickly generate low frequency correction $\mathcal{S}_{\ell}\bm{r}_\ell^{2h}$ in $k$-space, and then use ${ {F}^{H}}$ to convert corrected error from $k$-space to image space. Mathematically, we can formulate it as
\[\bm{e}^{2h}_\ell={ {F}^{H}}\mathcal{S}_{\ell}\bm{r}_\ell^{2h},\]
where the operation $\mathcal{S}_\ell$ is composed of four residual blocks (ResBlocks) \cite{He2016} with $p$ channels, two $3\times 3$ convolution layers with $p$ channels and a skip connection as shown in Fig.\ref{fig_architecture}(e). Obviously, the block ${ {F}^{H}}\mathcal{S}_{\ell}$ can be seen as an approximation to inverse operation $(\mathcal{T}^{2 h}_\ell)^{-1}$.
{ As we all know, solving inverse operator $(\mathcal{T}^{2 h}_\ell)^{-1}$ mathematically (e.g., conjugate gradient method) requires expensive computational costs. Although setting $\mathcal{S}_{\ell}$ to be learnable will increase the training complexity of the network, it is more convenient than solving the inverse operator mathematically in the testing stage. In addition, the amount of training data in this work is sufficient to effectively train such solution operator $\mathcal{S}_{\ell}$.} We also remark that Batch Normalization (BN) is not adopted because some recent papers showed that BN layer is more likely to yield undesirable representations when the network becomes deeper and more complex 
\cite{Wang2018,Zhang2018a}.

Next, we interpolate the correction $\bm{e}^{2h}_\ell$ on coarse level to the fine level by
\[\bm{e}_{\ell}^{h}=\mathcal{I}_\ell^{P}\bm{e}^{2h}_\ell,\]
where $\mathcal{I}_\ell^{P}$ is implemented by a learnable transpose convolution with stride 2 pixel to represent the prolongation operator $I_{2 h}^{h}$.

Using channel concatenation
to add contextual information $\bm{e}^{h}_\ell$ has been widely used
in traditional CNN architectures. 
According to {\bf Step 3}, we thus apply
a concatenate operation $\text{Concat}(\cdot,\cdot,\cdot)$ to produce the refined
correction module $\mathcal{C}_\ell$ as
\begin{equation*}
        \begin{split}
                \bm{m}_{\ell}&=\mathcal{C}_\ell(\bm{h}_{\ell},\bm{y},\mathcal{T},\mathcal{I}_{\ell}^{R},\mathcal{S}_{\ell},\mathcal{I}_{\ell}^{P},\beta_{\ell})\\
                &=\text{Concat}(\bm{h}_{\ell},\bm{e}_{\ell}^{h},\mathcal{M}_\ell^{\beta_\ell})=\text{Concat}(\bm{h}_{\ell},\mathcal{I}_{\ell}^{P}\bm{e}^{2h}_\ell,\mathcal{M}_\ell^{\beta_\ell}),
        \end{split}
        \label{eq9}
\end{equation*}
where the additional noise level map $\mathcal{M}_\ell^{\beta_\ell}$, which is filled with the output $\beta_\ell$ of condition module $\mathcal{CM}$ \cite{Zhang2020,You2021}, has the same size as $\bm{h}_{\ell}$ and can make network more flexible to train multi-sampling-ratio CS-MRI task.

\subsubsection{Geometric prior distillation module $\mathcal{P}_\ell$} We employ a geometric prior distillation module $\mathcal{P}_\ell$ \cite{Fan2021} to refine the compromised image structure and geometric features (\emph{see} Fig.\ref{fig_architecture}(d)).

{Actually, the geometric texture information in different sparse (geometric) domains can be represented by the partial derivatives $\mathcal{N}(\bm{x})=-\lambda\frac{\partial\mathcal{R}(\bm{x})}{\partial\bm{x}}$ of the geometric 
prior $\mathcal{R}(\bm{x})$ of MRI image $\bm{x}$. $\mathcal{N}(\bm{x}_{\ell})$ represents the geometric characteristics of $\bm{x}_{\ell}$ \cite{Zhang2018}, and $\mathcal{N}^k(\bm{m}_{\ell})$ extracts the $k$-order geometric characteristics of $\bm{m}_{\ell}$ \cite{Zhang2018}.  In this part, we hope that the geometric characteristics $\mathcal{N}(\bm{x}_{\ell})$ can be approximated by a linear combination of $\mathcal{N}(\bm{m}_{\ell}),...,\mathcal{N}^{k}(\bm{m}_{\ell}), R(\mathcal{N}^{k})$.}
Unfortunately, it is difficult to directly obtain the close-form solution $\bm{x}_{\ell}=(I-\mathcal{N})^{-1}(\bm{m}_{\ell})$ of the optimal condition \eqref{eq7} of the geometric prior distillation stage (P2). { It is well-known that the refined correction solution $\bm{m}_{\ell}$ can be employed to restore high-resolution $\bm{x}_{\ell}$ by optimizing the problem (P2) or solving the nonlinear equation $\bm{x}_{\ell}-\mathcal{N}(\bm{x}_{\ell})=\bm{m}_{\ell}$. Fortunately, if the operator $\mathcal{N}$ satisfies the contraction condition $\|\mathcal{N}\|<1$, $(I-\mathcal{N})^{-1}$ can be simplified by using series expansion as follows
\begin{equation}
\bm{x}_{\ell}=(I-\mathcal{N})^{-1}(\bm{m}_{\ell})=\left(\sum_{i=0}^{k}\mathcal{N}^{i}+R(\mathcal{N}^{k})\right)(\bm{m}_{\ell}),
        \label{eq11}
\end{equation}
which means 
\begin{equation*}
\bm{x}_{\ell}\in\text{span}(\bm{m}_{\ell},\mathcal{N}(\bm{m}_{\ell}),...,\mathcal{N}^{k}(\bm{m}_{\ell}), R(\mathcal{N}^{k})),
\end{equation*}
where $R(\mathcal{N}^{k})$ is the remainder and can be approximated by $\mathcal{N}^{k+1}(\bm{m}_{\ell})$.
}

A more flexible representation of the non-linear operation $\mathcal{N}^j$ in (\ref{eq11}) could be approximated by CNN block $\mathcal{N}^j_{\mathcal{K},\ell}$ with many embedded convolution blocks and ReLU layers, where $\mathcal{G}_{\ell,k}$ is designed to replace $\mathcal{N}$ for learning multi-scale geometric priors of $\bm{x}_{\ell}$.
Finally, a $1\times1$ convolution $\mathcal{W}_{\ell}=\left\{\mathcal{W}_{\ell,j}\right\}_{j=0}^{k}$
is used to fuse features $\mathcal{F}_{\ell,k}$, i.e.
\begin{align}
        \bm{x}_{\ell}&=\mathcal{P}_\ell(\bm{m}_\ell)=\mathcal{K}_{\ell,0}\ast\bm{m}_{\ell}+\sum_{j=1}^{k} \mathcal{N}^j_{\mathcal{K},\ell}\left(\bm{m}_{\ell}\right) \notag\\
        &=\mathcal{K}_{\ell,0}\ast\bm{m}_{\ell}+\sum_{j=1}^{k}\left(\mathcal{K}_{\ell,j}\ast\mathcal{B}^{j-1}_{\ell}(\mathcal{A}_{\ell}(\bm{m}_{\ell}))\right) \label{eq11A}
        \\
        &=\mathcal{W}_{\ell}\ast\mathcal{F}_{\ell,k} \notag
        \\
        &=\mathcal{W}_{\ell}\ast\text{Concat}\left(\mathcal{F}_{\ell,k-1},\mathcal{G}_{\ell,k}*\mathcal{F}_{\ell,k-1})\right), \notag
\end{align}
where $\mathcal{A}_{\ell}$ and $\mathcal{B}_{\ell}$ are two learnable blocks with the $3\times3$ convolution $\bm{\kappa}$ and ReLU such that
$\mathcal{A}_{\ell}(\cdot):=\text{ReLU}(\bm{\kappa}(\cdot)):\mathbb{R}^{m\times n}\rightarrow \mathbb{R}^{m\times n\times p}$, $\mathcal{B}_{\ell}(\cdot):=\text{ReLU}(\bm{\kappa}(\cdot)):\mathbb{R}^{m\times n\times p}\rightarrow \mathbb{R}^{m\times n\times p}$, $\mathcal{B}_{\ell}^{k-1}=\underbrace{\mathcal{B}_{\ell}\circ\dots\circ\mathcal{B}_{\ell}}_{k-1}$, $\mathcal{G}_{\ell,k}=(\underbrace{\mathcal{A}_{\ell},\mathcal{B}_{\ell},\dots,\mathcal{B}_{\ell}}_{k})$, each $1 \times 1$ convolution $\mathcal{K}_{\ell,j}$ in  $\mathcal{K}_{\ell}=\left(\mathcal{K}_{\ell,0}, \mathcal{K}_{\ell,1}, \ldots, \mathcal{K}_{\ell,k}\right) $ is a linear combination of $\left\{\mathcal{W}_{\ell,j}\right\}_{j=0}^{k}$.  We  also refer the interested readers to \cite{Fan2021} for more details.

\subsubsection{Condition module $\mathcal{CM}$}\label{section_CM} The pre-relaxation reconstruction step (\ref{eq4}) involves choosing/tuning hyper-parameter $\eta_\ell$
that significantly affects model performance.
A suitable step-length $\eta_\ell$ can achieve a satisfactory reconstruction as long as the CS $k$-space sampling-ratio is fixed.
However, it is generally difficult to determine a suitable $\eta_\ell$ without which the reconstructed image can be unsatisfactory,
especially when switching to other medical data with multi-sampling-ratio or other sampling masks. In general,  manual  tuning to determine optimal task-specific hyper-parameters
require high computational costs and often lead  to sub-optimal results.

To make multi-sampling-ratio tasks more flexible in our unifying model and tackle the inefficiency of hyper-parameter tuning (which
includes both adaptive model parameters and hyper-parameters), we learn the   step-length $\eta_{\ell}$ and noise level $\beta_\ell$ in our condition module $\mathcal{CM}$. We remark that CS sampling-ratio $\alpha$ influences the degree of ill-posedness, so it is suitable as the input of condition module $\mathcal{CM}$, resulting in the commonly used parameter prediction module \cite{You2021}, i.e.,
\begin{equation}
        [\bm{\eta},\bm{\beta}]=\mathcal{CM}(\alpha),
        \label{eq12}
\end{equation}
where $\bm{\eta}=[\eta_1,\eta_2,...,\eta_{n_s}]$ and $\bm{\beta}=[\beta_1,\beta_2,\dots,\beta_{ n_s}]$. { $\mathcal{CM}$ module is composed of three hidden fully connected layers with $p$ neurons, 
ReLU is employed as the first two activation functions. Since the pre-relaxation module $\mathcal{D}_\ell$ will not work when the step-length $\eta_\ell=0$, so softplus \cite{Glorot2011} is used as the last activation function.} 

\subsubsection{Loss function}
For the training data pairs $\left\{\left(\bm{y}^{i}, \bm{x}^{i}\right)\right\}_{i=1}^{n_t}$ (with $n_t$   the total number of samples), under-sampled $k$-space data $\bm{y}$ and  initialization $\bm{x}_{0}$ are used as inputs, the final output $\bm{x}_{n_s}$ is obtained by the proposed framework. The loss function is commonly employed to seek the real target image $\bm{x}^*$ by minimizing the distance measure between $\bm{x}_{\scriptscriptstyle n_s}$ and $\bm{x}$. Here we adopt $\ell_1$-loss rather than $\ell_2$-loss which is insufficient to capture perceptually relevant components (e.g., high-frequency geometric details) \cite{KJiang2018} to enlarge the original loss.

Inspired by deep supervision technique \cite{Lee2015}, we add an auxiliary constraint on a network branch to supervise the trunk network   to make the optimization process more flexible. The total loss is defined as follows
\begin{equation}
        \mathcal{L}=\frac{1}{n_t mn} \sum_{i=1}^{n_t}\left(\left\|\bm{x}^{i}_{\scriptscriptstyle\overline{ n_s}}-\bm{x}^{i}\right\|_{1}+\left\|\bm{x}^i_{\scriptscriptstyle n_s}-\bm{x}^{i}\right\|_{1}\right),
        \label{eq13}
\end{equation}
where $\overline{ n_s}=\frac{ n_s+1}{2}$, $mn$ is the size of $\bm{x}^{i}$ from \eqref{eq11A}, and $ n_s$ is the total stage number of the proposed framework. $\bm{x}^{i}_{\scriptscriptstyle\overline{ n_s}}$ and $\bm{x}^i_{\scriptscriptstyle n_s}$ are two outputs of the sample $\bm{x}^{i}$ on stages $\overline{ n_s}$ and $n_s$, respectively.

\subsubsection{Parameters and Initialization}
The three modules in stage $\ell$ of the proposed framework strictly inplement our correction-distillation updates (\eqref{eq4} to \eqref{eq7}). The learnable parameter set $\Theta=\left\{\eta_{\ell},\beta_{\ell}, \mathcal{A}_{\ell},\mathcal{B}_{\ell},\mathcal{K}_{\ell},\mathcal{I}_{\ell}^{R},\mathcal{S}_{\ell},\mathcal{I}_{\ell}^{P}\right\}_{\ell=1}^{ n_s}$ consists of step-length $\eta_{\ell}$, noise { level} $\beta_{\ell}$, CNN blocks $\mathcal{A}_{\ell}$ and $\mathcal{B}_{\ell}$ with $3\times3$ convolution and ReLU, $1 \times 1$ convolution $\mathcal{K}_{\ell}$, restriction operator $\mathcal{I}_{\ell}^{R}$, solution operator $\mathcal{S}_{\ell}$ and prolongation operator $\mathcal{I}_{\ell}^{P}$. All these parameters are learned as neural network parameters by minimizing the loss (\ref{eq13}).

Similar to traditional model-based method, the proposed framework also requires an initial input $\bm{x}_{0}={ \mathcal{T}^{H}}\bm{y}$.
The convolution network is initialized with Kaiming Initialization \cite{He2015}. The model parameters $\{p, k, n_{s}\}$ are initialized as $\{32, 8, 13\}$ respectively.

\section{Experiments and results}
In this section, we verify the advantages of the proposed CGPD-CSNet in comparing with the state-of-the-art methods through various experiments. Peak Signal to Noise Ratio (PSNR) and Structural Similarity Index Measure (SSIM) are employed to evaluate their performances, the values (M-PSNR and M-SSIM)
are all obtained by averaging on tested data.

\begin{table*}
        \centering
        \caption{Quantitative assessment with M-PSNR values of different stage number $ n_s$ and different geometric distillation depths $k$ using Cartesian sampling brain dataset with different CS ratios.}
        \label{table_stage_number}
        \setlength{\tabcolsep}{2.mm}{
                \begin{tabular*}{\hsize}{@{}@{\extracolsep{\fill}}lcccccc ccccccc@{}}      
                \toprule[1.5pt]
                \multirow{2}*{\textbf{CS Ratio}}&\multicolumn{6}{c}{\textbf{Stage number $ n_s$}}   &\multirow{2}*{ }&\multicolumn{6}{c}{\textbf{Geometric distillation depths $k$
                }}\\
                \cmidrule[0.7pt](lr){2-7} \cmidrule[0.7pt](lr){9-14}
                & 5 & 7 & 9 & 11 & \textbf{13} & 15  &   & 4 & 5 & 6 & 7 & \textbf{8} & 9\\
                \midrule[0.8pt]
                10\% & 32.54 & 33.59 & 33.80 & 34.16 & 34.23 & \textbf{34.25}    &  & 33.69 & 33.91 & 34.02 & 34.22 & 34.23 & \textbf{34.37}\\
                20\% & 37.43 & 38.31 & 38.61 & 38.88 & \textbf{39.02} & 39.00     &  & 38.46 & 38.65 & 38.75 & 38.93 & \textbf{39.02} & 39.02\\
                30\% & 39.56 & 40.46 & 40.76 & 41.01 & \textbf{41.16} & 41.15     &  & 40.64 & 40.83 & 40.89 & 41.10 & \textbf{41.16} & 41.14\\
                40\% & 43.45 & 44.38 & 44.71 & 44.92 & 45.07 & \textbf{45.08}    &   & 44.71 & 44.84 & 44.80 & 44.99 & \textbf{45.07} & 45.01\\
                50\% & 45.99 & 46.72 & 47.05 & 47.19 & 47.31 & \textbf{47.38}    &   & 47.05 & 47.13 & 47.10 & 47.28 & \textbf{47.31} & 47.29\\
                \bottomrule[1.5pt]
\end{tabular*}}
\end{table*}

\subsection{Implementation Details}
{We evaluate the performance of the proposed and the popular methods on the widely used brain MR dataset \cite{Yang2016,Zhang2018} using three classical types of sampling masks \cite{Yang2020}. The brain MR dataset consists of 100 images for training and 50 images for testing, which are T1-weighted 2D images from health and Alzheimer's disease patients from different devices. The training data is augmented to 800 by conducting horizontal and vertical flipping. The brain MR datase is randomly selected from the MICCAI 2013 SATA Challenge \cite{NNA2015}, which consists of 35 training and 12 testing T1-weighted scans of 1 mm thick cortical slices from the OASIS project \cite{Marcus2007} with corresponding manually created reference labels for 14 deep brain structures.} 

Besides, we employ a cardiac dataset \cite{Zheng2019}, which is established based on \cite{Andreopoulos2008} and contains 4480 cardiac real-valued MR images from 33 patients, to further evaluate the performance of comparison methods. The first 30 patients' samples of 4180 MR images are set as training set while the last 3 patients' of 300 MR images are set as testing set. The training data is augmented by conducting horizontal and vertical flipping. {The cardiac dataset is provided by the Department of Diagnostic Imaging of the Hospital for Sick Children in Toronto, Canada \cite{Andreopoulos2008}. The images are scanned with a GE Genesis Signa MR scanner using the FIESTA scan protocol. Most of the subjects display a variety of heart abnormalities such as cardiomyopathy, aortic regurgitation, enlarged, etc.}

We employ Pytorch to implement the proposed framework. We use Adam optimization \cite{Kingma2014} with a learning rate of $0.0001$ and batch size $1$ to train network for 500 epochs to ensure convergence. In our unifying multi-sampling-ratio CS-MRI framework, the mixed CS sampling ratios are set as $\{10\%, 20\%, 30\%, 40\%, 50\%\}$ for Cartesian sampling mask and pseudo radial sampling mask, and are set as $\{5\%, 10\%, 20\%, 30\%, 40\%\}$ for 2D random sampling mask. All experiments are performed on a workstation with Intel Xeon CPU E5-2630 and Nvidia Tesla V100 GPU.

\subsection{Intra-Method Evaluation}
We first conduct four groups of experiments to investigate the role of different network components in our CGPD-CSNet on reconstruction performance, including stage number $ n_s$, geometric prior distillation depths $k$, as well as ablation study about the proposed correction module $\mathcal{C}_\ell$ and condition module $\mathcal{CM}$, different shared settings of CGPD-CSNet.

\subsubsection{Test of stage number $ n_s$}
To evaluate the effectiveness of different stage number on reconstruction performance, we tune the stage number $ n_s$ from 5 to 15 at 2 intervals. Using the CNN architectures with different stage number $ n_s$ to Cartesian sampling brain dataset with different CS sampling ratios, the M-PSNR values of the reconstructed results are summarized in Table \ref{table_stage_number}. We can observe that the reconstruction performance gradually improves with the increase of stage number $ n_s$ while the M-PSNR value becomes stable after $ n_s\geq13$. Based on this observation, the 13-stage configuration is a preferable setting to balance the reconstruction performance and computational costs, and we fix $ n_s=13$ throughout all the experiments.

\subsubsection{Test  of geometric distillation depths $k$}
To explore the relationship between different geometric prior distillation depths and reconstruction performance, we tune the geometric distillation depths $k$ from 4 to 9. The M-PSNR values of the reconstructed images on brain dataset using different geometric distillation depths and  different Cartesian CS sampling ratios are summarized in Table \ref{table_stage_number}. The reconstruction performance improves slowly after $k\geq8$. Considering the tradeoff between network complexity and reconstruction performance, we set geometric distillation depths $k=8$ in all configurations.

\subsubsection{Ablation study}
Next, we conduct a group of ablation studies to better evaluate the effectiveness of correction module $\mathcal{C}_\ell$ and condition module $\mathcal{CM}$ on the CS-MRI reconstruction performance. The comparisons are shown in Table \ref{table_ablation}.

\begin{table}
\centering
\caption{Quantitative assessment with M-PSNR values of different combinations of correction module $\mathcal{C}_\ell$ and condition module $\mathcal{CM}$ using Cartesian sampling mask on brain dataset. The best and second places are highlighted in Bold font and underline ones, respectively.}
\label{table_ablation}
\setlength{\tabcolsep}{1.5mm}{
        \begin{tabular}{ccccccccc}
                \toprule[1.5pt]
                \multirow{2}*{\textbf{Variant}} &\multirow{2}*{\textbf{$\mathcal{C}_\ell$}} &\multicolumn{2}{c}{\textbf{$\mathcal{CM}$}} & \multicolumn{5}{c}{\textbf{CS Ratio}}\\ \cmidrule[0.7pt](lr){3-9} 
                & & $\eta_\ell$& $\beta_\ell$ &10\%&20\%& 30\% & 40\% & 50\% \\ \midrule[0.8pt]
                (a)  & -&-&-& 32.53&38.32& 40.63&44.60&46.66 \\
                (b)  & -&+&+&32.61&38.40&40.69&44.80&47.16\\
                (c)  & +&-&-&33.54 &38.54&40.76&44.66&46.89\\
                (d)  & +&-&+&33.90 &38.63& 40.81&44.78&47.11\\
                (e)  & +&+&-&\underline{34.00} &\underline{39.00}&\underline{41.13}&\underline{45.04}&\underline{47.30}\\
                (f)  & +&+&+&\textbf{34.23}&\textbf{39.02}&\textbf{41.16}&\textbf{45.07}& \textbf{47.31} \\
                \bottomrule[1.5pt]
\end{tabular}}
\end{table}

\begin{table}
\centering
\caption{Quantitative assessment with M-PSNR values with different shared settings of proposed CGPD-CSNet using Cartesian sampling mask on brain dataset. Bold font in the table is the best of the variants.}
\label{table_shared}
\setlength{\tabcolsep}{0.8mm}{
        \begin{tabular}{ccccccc}
                \toprule[1.5pt]
                \multirow{2}*{\textbf{Variant}} &\multirow{2}*{\textbf{Shared setting}} & \multicolumn{5}{c}{\textbf{CS Ratio}}\\ \cmidrule[0.7pt](lr){3-7}
                &&10\%&20\%& 30\% & 40\% & 50\% \\ \midrule[0.8pt]
                (a)  &Shared $\mathcal{C}_\ell$, $\mathcal{P}_\ell$ & 32.84&38.31& 40.55&44.65&47.02 \\
                (b)  &Shared $\mathcal{P}_\ell$& 33.55&38.55&40.76&44.79&47.09\\
                (c)  & Shared $\mathcal{C}_\ell$ &33.56 &38.71&40.93&44.91&47.17\\
                (d)  &Unshared (default) &\textbf{34.23}&\textbf{39.02}&\textbf{41.16}&\textbf{45.07}& \textbf{47.31}\\
                \bottomrule[1.5pt]
\end{tabular}}
\end{table}

\begin{table*}
\centering
\newsavebox{\mybox}
\begin{lrbox}{\mybox}
        \setlength{\tabcolsep}{2mm}{
                \begin{tabular*}{\hsize}{@{}@{\extracolsep{\fill}}llcccccc@{}}       
                \toprule[1.5pt]
                \multirow{2}*{\textbf{Dataset}} &\multirow{2}*{\textbf{Method}} & \multicolumn{5}{c}{\textbf{CS Ratio}} &  \textbf{Time (s)} \\ \cmidrule[0.7pt](lr){3-7}
                
                & & 10\% & 20\% & 30\% & 40\% & 50\% & \textbf{CPU/GPU} \\
                \midrule[0.8pt]
                \multirow{11}*{\textbf{Brain}} &Zero-filling\cite{Bernstein2001} & 23.85/0.5743& 26.40/0.6611&   28.64/0.7249&   31.12/0.7863 &  32.93/0.8230 & 0.0019/---\\
                &TV\cite{Lustig2007} & 25.56/0.6685& 29.92/0.7939& 32.88/0.8582&    37.19/0.9285 & 39.91/0.9472 & 1.2330/---     \\
                &{SIDWT} & 25.19/0.6782& 28.96/0.7782& 31.87/0.8476& 36.50/0.9234   & 39.70/0.9493 & 18.033/---\\
                &{PBDW}\cite{Qu2012}  & 26.82/0.7355& 32.42/0.8661& 35.04/0.9116& 39.76/0.9599  & 42.52/0.9727   & 65.1656/---\\
                &{PANO}\cite{Qu2014}  &28.98/0.7897&34.64/0.8934& 36.79/0.9244& 41.54/0.9661 & 44.14/0.9733& 38.2577/---\\
                &{DC-CNN}\cite{Schlemper2018}  & 30.00/0.8232&35.73/0.9204& 38.48/0.9493&42.87/0.9765& 45.53/0.9842& ---/0.0161 \\
                &{ISTA-Net}\cite{Zhang2018} & 30.28/0.8256 &36.59/0.9334 &39.14/0.9566&43.93/0.9820&  46.77/0.9885&---/0.0189\\
                &{ISTA-Net+}\cite{Zhang2018}& 30.86/\underline{0.8455}&37.05/0.9356 &39.62/0.9597&44.24/0.9826& 46.78/0.9882 &{---/0.0200}\\
                &{FISTA-Net}\cite{Xiang2020} & \underline{31.06}/0.8444&    \underline{37.27}/\underline{0.9395} &  \underline{39.86}/\underline{0.9607} &   \underline{44.50}/\underline{0.9835} & \underline{46.99}/\underline{0.9889} &{---/0.0297}\\
                &{CGPD-CSNet} &\textbf{34.23}/\textbf{0.9126}&\textbf{39.02}/\textbf{0.9553}&\textbf{41.16}/\textbf{0.9685}&\textbf{45.07}/\textbf{0.9852}& \textbf{47.31}/\textbf{0.9897} &{---/0.0721}\\
                \midrule[0.8pt]
                \multirow{11}*{\textbf{Cardiac}} &Zero-filling\cite{Bernstein2001} & 20.29/0.5286&   22.99/0.6059&   25.50/0.6698&   28.05/0.7371&   29.97/0.7736 & 0.0017/---\\
                &TV\cite{Lustig2007} & 22.26/0.6139& 27.94/0.7786& 31.32/0.8535&    36.27/0.9274 & 38.87/0.9486 & 1.2446/---     \\
                &{SIDWT} & 21.56/0.6233& 26.23/0.7480& 29.62/0.8279& 35.02/0.9102   & 37.97/0.9416 & 21.1174/---\\
                &{PBDW}\cite{Qu2012}  & 23.73/0.7123& 30.46/0.8666& 33.18/0.9122& 38.58/0.9642  & 41.20/0.9761   & 69.8070/---\\
                &{PANO}\cite{Qu2014}  &25.94/0.7606&32.45/0.8809& 34.61/0.9155& 40.81/0.9697 & 43.07/0.9730& 39.0812/---\\
                &{DC-CNN}\cite{Schlemper2018}  & 26.31/0.7863&33.51/0.9153& 36.29/0.9484&42.56/0.9836& 45.25/0.9888& ---/0.0131 \\
                &{ISTA-Net}\cite{Zhang2018} & 27.81/0.8185&34.23/0.9299&36.70/0.9557&43.94/0.9892&  47.51/0.9941&---/0.0134\\
                &{ISTA-Net+}\cite{Zhang2018}& \underline{27.84}/0.8236&34.57/0.9339&37.20/0.9589&44.56/0.9903&  48.07/0.9946 &{---/0.0145}\\
                &{FISTA-Net}\cite{Xiang2020} & 27.74/\underline{0.8280}&     \underline{34.70}/\underline{0.9347}&  \underline{37.41}/\underline{0.9600} &  \underline{44.83}/\underline{0.9905} & \underline{48.37}/\underline{0.9949} &{---/0.0229}\\
                &{CGPD-CSNet} &\textbf{28.05/0.8516}&\textbf{35.32}/\textbf{0.9433}&\textbf{37.99}/\textbf{0.9649}&\textbf{45.47}/\textbf{0.9919}& \textbf{48.66}/\textbf{0.9952} &{---/0.0539}\\
                \bottomrule[1.5pt]
\end{tabular*}}
\end{lrbox}
\caption{Quantitative assessment with M-PSNR and M-SSIM values using Cartesian sampling mask with different CS ratios on brain and cardiac datasets. The best and second best results are highlighted in Bold font and underlined ones, respectively.}
{\usebox{\mybox}}\label{table_comparison}
\end{table*}

Using condition module $\mathcal{CM}$ with two learnable parameters, our method enjoys the flexibility of handling CS-MRI problems with different sampling ratios through a single model. Contrast to the single variant (a) trained with the same Cartesian sampling masks without condition module $\mathcal{CM}$, variant (b) with the proposed condition module $\mathcal{CM}$ consistently outperforms variant (a) across all five CS ratios and brings average 0.18 dB improvement. When the proposed correction module $\mathcal{C}_\ell$ is embedded in reconstruction model, variant (f) with condition module $\mathcal{CM}$ can further improve reconstruction performance than variant (c) without condition module $\mathcal{CM}$ and increases average $0.48$ dB. Furthermore, comparison between variants (a) and (c) can also demonstrate the advantage of the proposed correction module $\mathcal{C}_\ell$. Especially in low sampling rate $10\%$, reconstruction performance improves $1.01$ dB.

From variants (c) and (d), it is clear to descript that noise level $\beta_\ell$ learned from condition module $\mathcal{CM}$ can improve the M-PSNR value by $0.17$ dB. We can also observe from variants (c) and (e) that step-length $\eta_\ell$ learned from condition module $\mathcal{CM}$ can greatly boost the reconstruction performance across all five CS ratios and achieve average $0.42$ dB improvement. Compared with variant (a), our variant (f) combining with correction module $\mathcal{C}_\ell$ and condition module $\mathcal{CM}$ can achieve average $0.81$ dB improvement across all five sampling ratios and $1.7$ dB in low CS ratio $10\%$ especially. These comparisons adequately verify the effectiveness of the proposed correction module $\mathcal{C}_\ell$ and condition module $\mathcal{CM}$.

\subsubsection{Module sharing configurations}
To demonstrate the flexibility of the proposed framework that does not have to be the same network parameter configurations in different stages, we conduct several variants of CGPD-CSNet that have different shared settings among stages. Table \ref{table_shared} lists the M-PSNR values for different shared architectures in case of using Cartesian sampling mask on brain dataset. Note that the best M-PSNR scores are achieved when using the default unshared version (d), which is the most flexible with largest number of parameters. The variant (a) that shares both $\mathcal{C}_\ell$ and $\mathcal{P}_\ell$ in all stages is least flexible with smallest number of parameters and achieves the worst performance. Especially if only $\mathcal{P}_\ell$ or $\mathcal{C}_\ell$ is shared, the variants (b) and (c) increase average 0.27 dB and 0.38 dB over the variant (a). So we adopt the default unshared version (d) to perform the following experiments.

Actually, we attribute the superiority of our method to the following three factors. Firstly, our method has a correction module $\mathcal{C}_\ell$ which can compensate the low frequency of reconstruction error in $k$-space. Secondly, the additional noise level $\beta_\ell$ learned from condition module $\mathcal{CM}$ adds the anti-interference, and the geometric prior distillation module $\mathcal{P}_\ell$ distills the lost contextual details. Thirdly, the learnable step-length $\eta_\ell$ can be adapted and optimized according to different CS sampling ratios, such that multi-sampling-ratio CS-MRI can be jointly learnt through a single model.

\subsection{Comparison to the state-of-the-art Methods}
Since our proposed CGPD-CSNet framework  merges the advantages of traditional methods and modern deep learning methods,
it is necessary to compare with related methods from both categories on performances and hence  generalizability.

We compare our framework with the state-of-the-art methods including traditional methods (Zero-filling \cite{Bernstein2001}, TV\cite{Lustig2007}, { SIDWT}\footnote{https://github.com/ricedsp/rwt}), patch-based methods (PBDW \cite{Qu2012}, PANO \cite{Qu2014}), 
and
deep unfolding methods (DC-CNN \cite{Schlemper2018}, ISTA-Net \cite{Zhang2018}, ISTA-Net+ \cite{Zhang2018}, FISTA-Net \cite{Xiang2020}. Following \cite{Zhang2018,Fan2021}, the stage number of ISTA-Net, ISTA-Net+ and FISTA-Net is configured as 11. Following \cite{Schlemper2018}, the stage number of DC-CNN is configured as 5. Cartesian sampling mask which is widely used in clinic \cite{Lustig2007,Qu2012} is employed to acquire under-sampling $k$-space data. Apart from other comparison methods that have to train different model according to every CS sampling ratio, we cast the training process for five different CS sampling ratios as a unified joint learning throughout all the experiments.

\begin{figure*}
\centering
\includegraphics[width=0.98\textwidth]{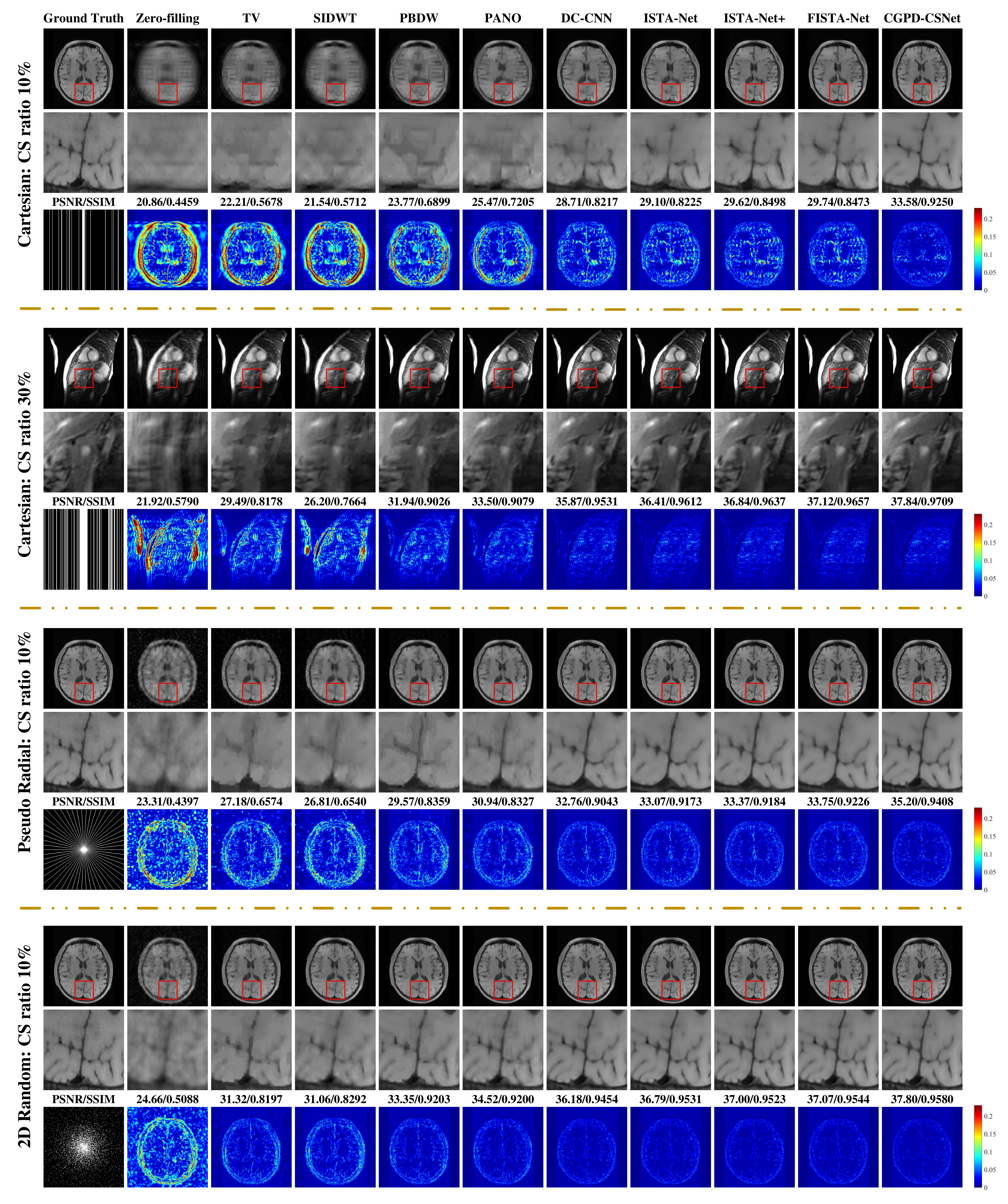}
\caption{{ Qualitative comparisons of our method with other popular methods using three sampling masks on brain and cardiac datasets.}}
\label{fig_results}
\end{figure*}

\begin{figure*}
\centering
\includegraphics[width=0.65\textwidth]{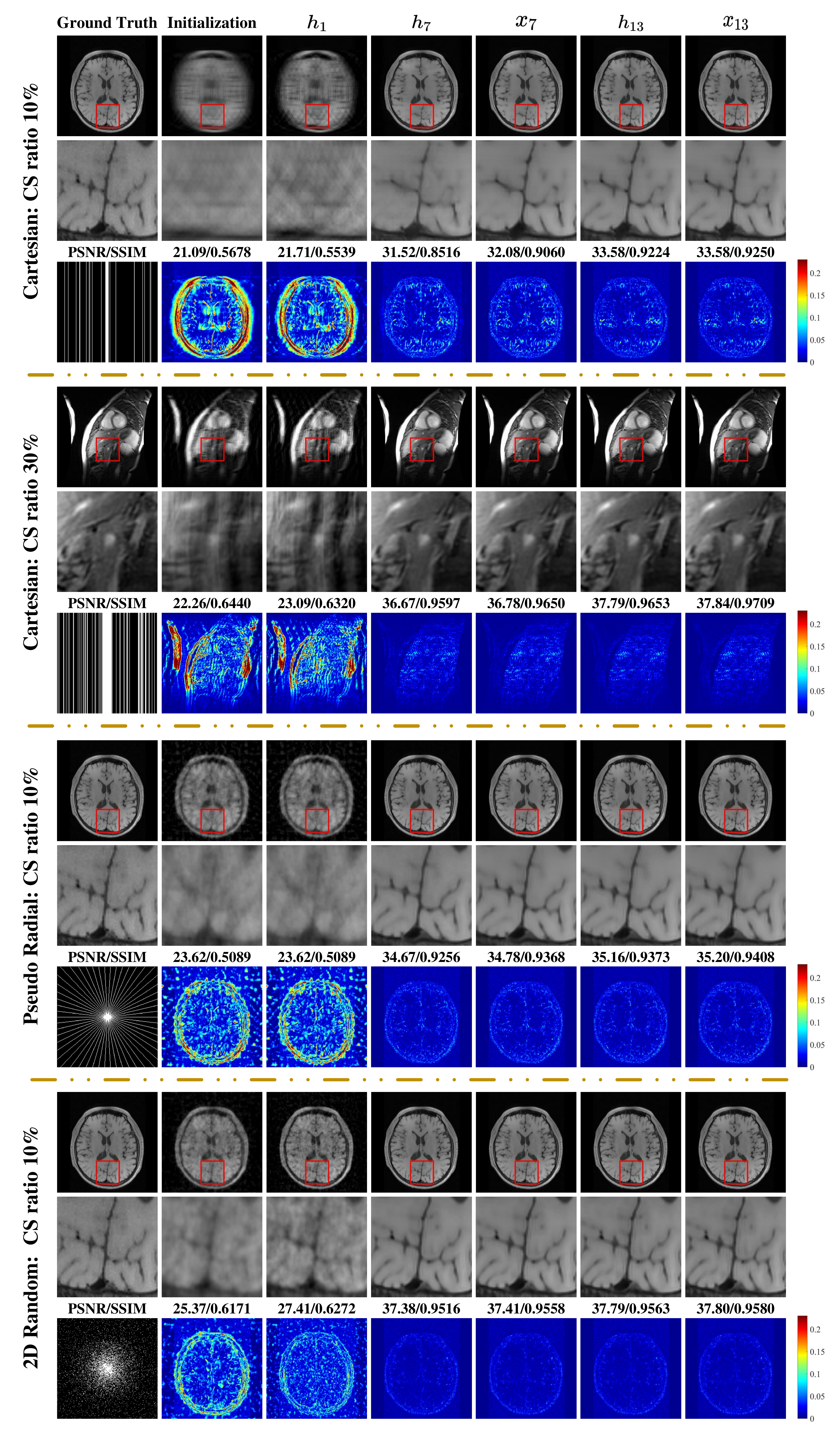}
\caption{{Stage representations of the proposed method using three sampling masks on brain and cardiac datasets.}}
\label{fig_iteration_image}
\end{figure*}

\begin{figure}[htbp]
        \centering
        \includegraphics[width=\columnwidth]{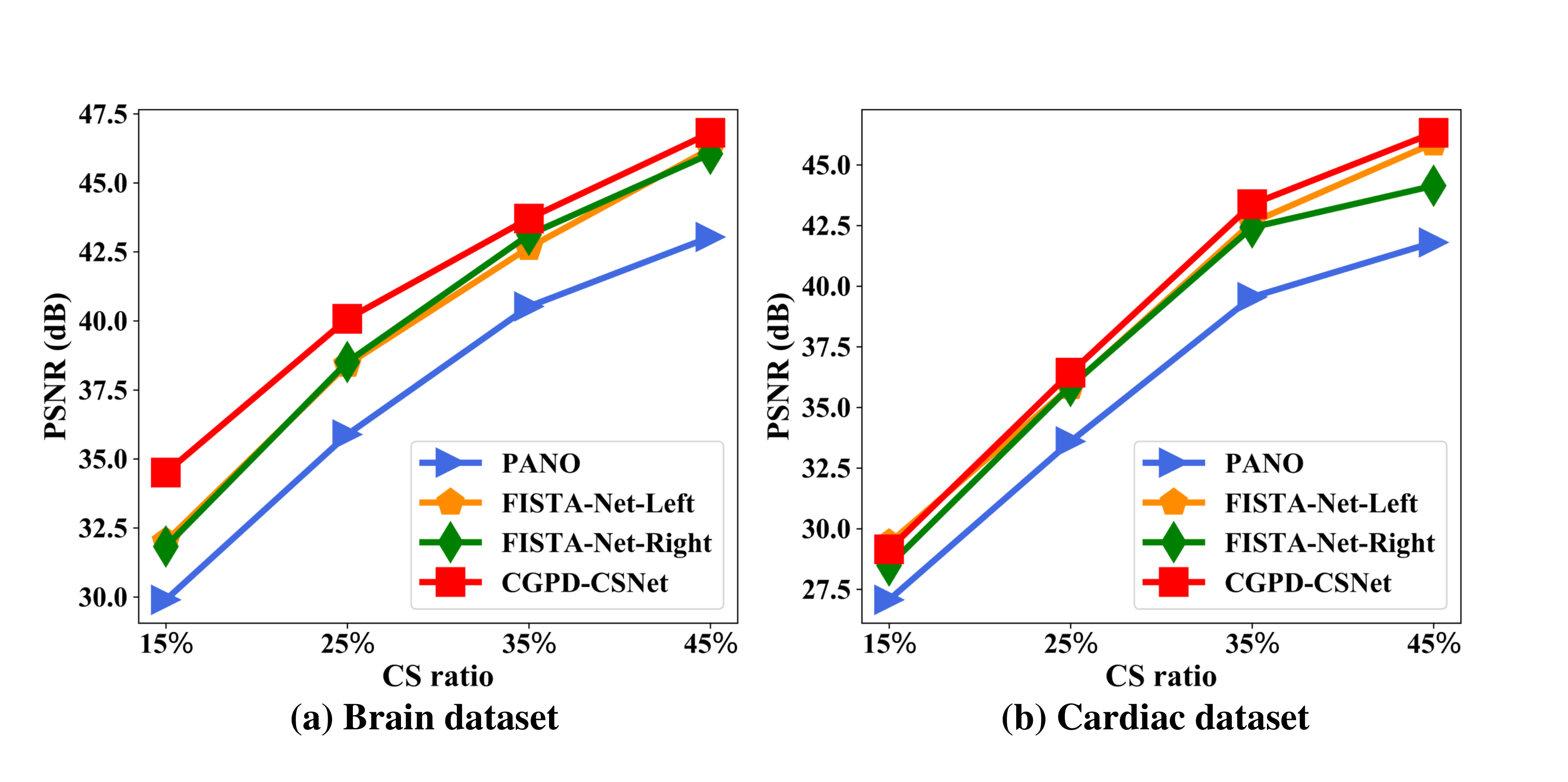}
        \caption{Performance comparisons on Cartesian under-sampled k-space datasets with
                untrained CS ratios for brain dataset (a) and cardiac dataset (b). "-Left" denotes the reconstruction of untrained CS ratio 15\%, 25\%, 35\%, 45\% sampled data with trained models of CS ratios 10\%, 20\%, 30\%, 40\%, respectively. "-Right" denotes the reconstruction of untrained CS ratio 15\%, 25\%, 35\%, 45\% sampled data with trained models of CS ratios 20\%, 30\%, 40\%, 50\%, respectively.}
        \label{fig_unsampled_compare}
\end{figure}


\subsubsection{Quantitative evaluation}
The M-PSNR/M-SSIM scores of the proposed and comparison methods using Cartesian sampling mask with different CS ratios on brain and cardiac datasets are listed in Table \ref{table_comparison}.
As a result, a significant increase of all M-PSNR/M-SSIM scores compared to the other methods across all CS ratios can be observed. In particular, our method achieves 3.18 dB improvement {(10.22\%)} at low CS ratio 10\% and average 1.42 dB improvement {(3.55\%)} over other comparison methods on brain dataset. And our method achieves average 0.47 dB improvement {(1.21\%)} over other comparison methods on cardiac dataset. Thanks to Nesterov acceleration and learning nonlinear transformations with deeper convolutional layers, FISTA-Net outperforms ISTA-Net+ across all CS ratios. The reconstruction results of all deep unfolding methods are better than traditional CS-MRI methods and patch-based methods. In addition, Table \ref{table_comparison} lists the average testing time (in seconds) of reconstructing a 256 $\times$ 256 image for all methods. Our method achieve consistently better reconstruction performance while maintaining a fast reconstruction real-time speed.

\begin{table*}
\centering
\begin{lrbox}{\mybox}
\setlength{\tabcolsep}{1.2mm}{
        \begin{tabular*}{\hsize}{@{}@{\extracolsep{\fill}}lcccccccccccc@{}}        
        \toprule[1.5pt]
        \multirow{2}*{\textbf{Method}}&\multicolumn{1}{c}{} & \multicolumn{5}{c}{\textbf{Pseudo radial CS Ratio $\alpha$}}&\multicolumn{1}{c}{}&\multicolumn{5}{c}{\textbf{2D random CS Ratio $\alpha$}} \\ \cmidrule[0.7pt](lr){3-7} \cmidrule[0.7pt](lr){9-13}
        &  & 10\% & 20\% & 30\% & 40\% & 50\% & & 5\%& 10\% & 20\% & 30\% & 40\% \\
        \midrule[0.8pt]
        Zero-filling\cite{Bernstein2001}& & 26.64& 30.28&  32.89&  35.01 & 36.92  &   & 25.98& 27.72&  29.34&  31.12&  33.01\\
        TV\cite{Lustig2007}& & 30.83 & 35.16 & 38.03 &  40.13 & 41.94  &  & 31.99& 34.34& 36.94& 39.34 & 41.58\\
        {SIDWT}& & 30.88 & 35.66 & 38.67 & 40.79& 42.57   &  & 31.79& 34.43& 37.36& 39.95 & 42.23\\
        {PBDW}\cite{Qu2012}&  & 32.48 & 36.54 & 38.90& 40.67& 42.30   &  & 33.34&35.67& 38.48& 40.83 & 42.94\\
        {PANO}\cite{Qu2014}&  &33.61 &37.02 & 39.28 & 41.05& 42.72   &  &34.31&36.62& 39.57& 42.04& 42.94\\
        {{ADMM-Net}\cite{Yang2016}}&  &{---} &{37.17} & {39.84} & {41.56}& {43.00}   &  &{---}&{---}& {---}&{---}&{---}\\
        {DC-CNN}\cite{Schlemper2018}&  & 34.30&38.43& 40.74&42.51 & 44.07 &  & 35.01&37.65& 40.80&43.34 & 45.56\\
        {ISTA-Net}\cite{Zhang2018}& & 34.69 &38.75 &41.00&42.26& 44.25  & & 35.59&38.23 &41.44&43.99& 46.21\\
        {ISTA-Net+}\cite{Zhang2018}&  & 34.83&38.75 &40.99&42.64& 44.22   &  & \underline{35.92}&\underline{38.41} &41.66&44.18& 46.51\\
        {FISTA-Net}\cite{Xiang2020}& & \underline{35.09}&\underline{39.05}& \underline{41.20} & \underline{42.85}& \underline{44.42}    &  & 35.92&38.36& \underline{41.74} & \underline{44.27}& \underline{46.69}\\
        {CGPD-CSNet}& &\textbf{36.22}&\textbf{39.54}&\textbf{41.52}&\textbf{43.08}& \textbf{44.58}  &  &\textbf{36.72}&\textbf{39.08}&\textbf{42.10}&\textbf{44.52}& \textbf{46.84}\\
        \bottomrule[1.5pt]
\end{tabular*}}
\end{lrbox}
\caption{Quantitative assessment with M-PSNR values using pseudo radial and 2D random sampling mask on brain dataset with different CS ratios for CS-MRI. The best and second places are highlighted in Bold font and underlined ones, respectively.}
{\usebox{\mybox}}\label{table_comparison_mask}
\end{table*}

\subsubsection{Qualitative evaluation}
The visual comparisons of all methods using Cartesian sampling mask with CS ratio 10\% on brain CS-MRI and CS ratio 30\% on cardiac CS-MRI are shown in Fig.\ref{fig_results}. We can observe that our method is able to restore more details (e.g., sharper edges, texture) which yield much better visual quality and achieve much higher PSNR scores than the comparison methods.


It is interesting and meaningful to qualitatively analyze the presentations of the pre-relaxation module $\mathcal{D}_\ell$ and the geometric prior distillation module $\mathcal{P}_\ell$ in different stages. Fig.\ref{fig_iteration_image} shows the intermediate visualizations of the proposed method at different stages by using Cartesian sampling mask with CS ratio 10\% on brain dataset and CS ratio 30\% on cardiac dataset. We can observe that the details of intermediate results become gradually more complete and that the reconstruction performance become better as iterations increase. It shows that both modules $\mathcal{D}_\ell$ and $\mathcal{P}_\ell$ can facilitate each other for alternating artifact removal and details recovery. These performances are also due to the deep supervision loss, which ensures that the reconstruction results can be steadily improved even with a large number of iterations.

In summary, from the perspectives of visualization and objective
evaluations, our method can achieve a prominent improvement across all CS ratios on different datasets by using widely used Cartesian sampling mask for clinical $k$-space data acquisition. More importantly, such excellent performance can be achieved by reconstructing multiple CS ratios tasks through a single model. Apparently, it can be attributed to the decoupling of prior term and the bonus of two-grid-cycle architecture that quickly generate the low frequency correction in $k$-space.

{\section{Discussions}
MRI is widely used for clinical diagnosis, while high-resolution MRI requires a long scan time and has the risk of motion-related artifacts in the reconstructed image. CS-MRI has been proposed to reconstruct image from the sparse characteristics of signals and the reconstructed image's quality is relatively high. Deep unfolding learnable framework inheriting the merits of model-based and DL-based CS-MRI methods has not only sufficient theoretical support but also good performance \cite{Yang2020,Wang2021,Aghabiglou2022}. However, there are still two main shortcomings: 1) the derivation from mathematical theory to network design is not always natural enough for these existing deep unfolding methods. The proximal-point sub-problem has not well been analyzed and explained;
2) most of existing CS-MRI methods are not flexible enough to handle multi-sampling-ratio reconstruction assignments.
        
We transform the proximal-point sub-problem to distill features of different geometric characteristic domains, and the derivation from mathematical theory to network design is more natural than the existing deep unfolding methods. Furthermore, the proposed framework can handle multi-sampling-ratio CS-MRI tasks in a single model to avoid expensive training cost and large storage, while the existing CS-MRI methods have to train an independent model for each CS ratio. We also propose a multi-grid inspired unfolding correction-distillation scheme which can not only incorporate frequency-domain information to compensate for low-frequency error in $k$-space, but also learn geometric priors of MR image by adding a geometric prior distillation module. 

{To achieve comprehensive evaluations, we conduct four groups of experiments to analyze the sensitivity of the proposed framework, including CS-MRI with untrained CS sampling ratio $\alpha$, CS-MRI with different sampling masks (pseudo radial sampling mask and 2D random sampling mask), different initialization $\bm{x}_0$ ($\bm{x}_0=\mathcal{T}^{H}\bm{y}$ and $\bm{x}_0=0$), and CS-MRI with Gaussian noise.}

\subsection{Sensitivity analysis of untrained CS ratios}
Fig.\ref{fig_unsampled_compare} depicts the test comparisons on Cartesian under-sampled $k$-space datasets with untrained CS ratios for brain and cardiac MR images. Here all comparison methods
trained five models according to five CS ratios 10\%, 20\%, 30\%, 40\% and 50\%, respectively, while the proposed unifying framework jointly train a single model on the above five CS ratios. Compared with other methods, our unifying framework improves at least 2.49 dB, 1.56 dB, 0.59 dB, and 0.58 dB for CS sampling ratio 15\% ,25\%, 35\%, 45\% on brain dataset respectively, with an average 1.30 dB improvement. And our unifying framework achieves 0.55 dB, 0.76 dB, 0.43 dB improvement for CS ratio 25\%, 35\%, 45\% on cardiac dataset. It is worth noting that our method can use the single multi-sampling-ratio model to reconstruct more satisfactory result from sampling data with any untrained CS ratio than other comparison methods.

\subsection{Sensitivity analysis of different sampling masks}
Table \ref{table_comparison_mask} lists the M-PSNR values of various methods by using pseudo radial sampling mask and 2D random sampling mask with different CS sampling ratios on brain dataset. We can see that the proposed method achieves optimal reconstruction results across all CS ratios.
In general, it can achieve average 0.47 dB improvement {(1.15\%)} using pseudo radial sampling mask and average 0.45 dB improvement {(1.08\%)} using 2D random sampling mask. Furthermore, we can observe that our method achieves much better visual quality and higher PSNR than other comparison methods in Fig.\ref{fig_results},
and that the details of intermediate results become gradually better as iterations increase in Fig.\ref{fig_iteration_image}
using pseudo radial sampling mask with CS ratio 10\% and 2D random sampling mask with CS ratio 10\% on brain dataset.

{\subsection{Sensitivity analysis of different initial inputs}
In this part, we conduct experiments with two typical initialization schemes, which are denoted by $\bm{x}_0=\mathcal{T}^{H}\bm{y}$ and $\bm{x}_0=0$ respectively, to evaluate the influence of initialization. Table \ref{table_initial} lists the quantitative assessment of different initial inputs $\bm{x}_0$ with M-PSNR values using Cartesian sampling mask and different CS ratios on brain dataset. We can find that the reconstruction performance of $\bm{x}_0=0$ is evidently worse than $\bm{x}_0=\mathcal{T}^{H}\bm{y}$ at a low CS ratio (10\%). Due to the efficient correction of multigrid technique in multiple resolutions, all two initializations achieve similar and good performance at the other four CS ratios. It demonstrates that the initialization we adopt is suitable.
\begin{table}[h]
        \centering
        \setlength\tabcolsep{6.5pt}
        \caption{Quantitative assessment of different initial inputs $\bm{x}_0$ with M-PSNR values using Cartesian sampling mask with different CS ratios on brain dataset.}
        \label{table_initial}
                \begin{tabular}{cccccc}
                        \toprule[1.5pt]
                        \multirow{2}*{\textbf{Initialization}} & \multicolumn{5}{c}{\textbf{CS Ratio}}  \\ \cmidrule[0.7pt](lr){2-6}
                        &  10\% & 20\% & 30\% & 40\% & 50\%  \\
                        \midrule[0.8pt]
                        $\bm{x}_0=\mathcal{T}^{H}\bm{y}$ &34.23&39.02&41.16&45.07&47.31\\
                        $\bm{x}_0=0$ &34.09&39.00&41.15&45.06& 47.31\\
                        \bottomrule[1.5pt]
        \end{tabular}
\end{table}
}

{\subsection{Sensitivity analysis of the noise degradation}
To show the robustness of the proposed method against image with the Gaussian noise degradation, we test the under-sampled $k$-space data $y$ with Gaussian noise using the trained model and untrained model. Table \ref{table_noise_psnr} lists the quantitative assessment of Gaussian noise degradation (mean value is 0 and the standard deviation is 0.1) with M-PSNR values using Cartesian sampling mask and different CS ratios on brain dataset. Thanks to the proposed two-grid-cycle correction architecture, we can see that our method is robust to gaussian noise regardless of whether it is trained. The reconstruction performance of the under-sampled $k$-space data with Gaussian noise will not drop too much.
\begin{table}[h]
        \centering
        \setlength\tabcolsep{4pt}
        \caption{Quantitative assessment of Gaussian noise with M-PSNR values using Cartesian sampling mask with different CS ratios on brain dataset.}
        \label{table_noise_psnr}
                \begin{tabular}{cccccc}
                        \toprule[1.5pt]
                        \multirow{2}*{\textbf{Noise}} & \multicolumn{5}{c}{\textbf{CS Ratio}}  \\ \cmidrule[0.7pt](lr){2-6}
                        &  10\% & 20\% & 30\% & 40\% & 50\%  \\
                        \midrule[0.8pt]
                        None &34.23&39.02&41.16&45.07&47.31\\
                        Gaussian (untrained) &34.17&38.89&40.98&44.68&46.78\\
                        Gaussian (trained) &34.23&38.88&41.02&44.82     &46.98\\
                        \bottomrule[1.5pt]
        \end{tabular}
\end{table}
}

Extensive numerical experiments also show that our framework outperforms state-of-the-art methods in terms of visualizations and quantitative evaluations on flexibility and stability. Our method achieves average 1.42 dB improvement on brain dataset and average 0.47 dB improvement on cardiac dataset over other popular methods using Cartesian sampling mask. It can achieve average 0.47 dB improvement using pseudo radial sampling mask and average 0.45 dB improvement using 2D random sampling mask on brain dataset. Otherwise, it is worth noting that our method can use the single multi-sampling-ratio model to reconstruct more satisfactory result from sampling data with any untrained CS ratio than other comparison methods. Therefore, our architecture has good generalizability for image reconstruction due to the two-scale correction-distillation, and can be potentially applied to other inverse problems in imaging. 

}

\section{Conclusions}
In this study, we proposed a novel deep unfolding unified framework that can deliver generalizability by flexibly handling multi-sampling-ratio CS-MRI through a single model. Inspired by efficient correction of multigrid technique, we start from classical CS-MRI optimization problem and transform it into a two-grid-cycle correction architecture which consists of pre-relaxation, correction and geometric prior distillation. The proposed method inherits the merits of model-based and DL-based CS-MRI methods, has sufficient theoretical support and also good performance. It can give us a new perspective to design the explainable network. Extensive experiments demonstrate that the proposed framework outperforms other state-of-the-art methods in terms of qualitative and quantitative evaluations.

\bibliographystyle{IEEEtran} 
\bibliography{CGPD_ref_R1}

\begin{thebibliography}{10}
\providecommand{\url}[1]{#1}
\csname url@samestyle\endcsname
\providecommand{\newblock}{\relax}
\providecommand{\bibinfo}[2]{#2}
\providecommand{\BIBentrySTDinterwordspacing}{\spaceskip=0pt\relax}
\providecommand{\BIBentryALTinterwordstretchfactor}{4}
\providecommand{\BIBentryALTinterwordspacing}{\spaceskip=\fontdimen2\font plus
\BIBentryALTinterwordstretchfactor\fontdimen3\font minus
  \fontdimen4\font\relax}
\providecommand{\BIBforeignlanguage}[2]{{%
\expandafter\ifx\csname l@#1\endcsname\relax
\typeout{** WARNING: IEEEtran.bst: No hyphenation pattern has been}%
\typeout{** loaded for the language `#1'. Using the pattern for}%
\typeout{** the default language instead.}%
\else
\language=\csname l@#1\endcsname
\fi
#2}}
\providecommand{\BIBdecl}{\relax}
\BIBdecl

\bibitem{Song2016}
L.~Song, J.~Zhang, and Q.~Wang, ``{MRI} reconstruction based on three
  regularizations: Total variation and two wavelets,'' \emph{Biomedical Signal
  Processing and Control}, vol.~30, pp. 64--69, Sep. 2016.

\bibitem{KYCui2021a}
K.~Cui, ``Dynamic mri reconstruction via weighted tensor nuclear norm
  regularizer,'' \emph{IEEE J. Biomed. Health Informat.}, vol.~25, no.~8, pp.
  3052--3060, 2021.

\bibitem{Zhan2016}
Z.~Zhan, J.-F. Cai, D.~Guo, Y.~Liu, Z.~Chen, and X.~Qu, ``Fast multiclass
  dictionaries learning with geometrical directions in {MRI} reconstruction,''
  \emph{IEEE Trans. Biomed. Eng.}, vol.~63, no.~9, pp. 1850--1861, Nov. 2016.

\bibitem{Lai2016}
Z.~Lai, X.~Qu, Y.~Liu, D.~Guo, J.~Ye, Z.~Zhan, and Z.~Chen, ``Image
  reconstruction of compressed sensing {MRI} using graph-based redundant
  wavelet transform,'' \emph{Med. Image Anal.}, vol.~27, pp. 93--104, Jan.
  2016.

\bibitem{Lingala2013}
S.~G. Lingala and M.~Jacob, ``Blind compressive sensing dynamic {MRI},''
  \emph{IEEE Trans. Med. Imag.}, vol.~32, no.~6, pp. 1132--1145, Jun. 2013.

\bibitem{Block2007}
K.~T. Block, M.~Uecker, and J.~Frahm, ``Undersampled radial {MRI} with multiple
  coils. iterative image reconstruction using a total variation constraint,''
  \emph{Magn Reson Med.}, vol.~57, no.~6, pp. 1086--1098, Jun. 2007.

\bibitem{Qu2012}
X.~Qu, D.~Guo, B.~Ning, Y.~Hou, Y.~Lin, S.~Cai, and Z.~Chen, ``Undersampled
  {MRI} reconstruction with patch-based directional wavelets,'' \emph{Magn.
  Resonance Imag.}, vol.~30, no.~7, pp. 964--977, Sep. 2012.

\bibitem{Qu2014}
X.~Qu, Y.~Hou, F.~Lam, D.~Guo, J.~Zhong, and Z.~Chen, ``Magnetic resonance
  image reconstruction from undersampled measurements using a patch-based
  nonlocal operator,'' \emph{Med. Image Anal.}, vol.~18, no.~6, pp. 843--856,
  Aug. 2014.

\bibitem{Metzler2016}
C.~A. Metzler, A.~Maleki, and R.~G. Baraniuk, ``From denoising to compressed
  sensing,'' \emph{IEEE Trans. Inf. Theory}, vol.~62, no.~9, pp. 5117--5144,
  Apr. 2016.

\bibitem{Beck2009}
A.~Beck and M.~Teboulle, ``A fast iterative shrinkage-thresholding algorithm
  for linear inverse problems,'' \emph{SIAM J. Imag. Sci.}, vol.~2, no.~1, pp.
  183--202, Jan. 2009.

\bibitem{Chambolle2010}
A.~Chambolle and T.~Pock, ``A first-order primal-dual algorithm for convex
  problems with~applications to imaging,'' \emph{J. Math. Imaging and Vis.},
  vol.~40, no.~1, pp. 120--145, Dec. 2010.

\bibitem{Boyd2010}
S.~Boyd, ``Distributed optimization and statistical learning via the
  alternating direction method of multipliers,'' \emph{Found. Trends Mach.
  Learn.}, vol.~3, no.~1, pp. 1--122, 2010.

\bibitem{Datta2022}
S.~Datta, S.~Dandapat, and B.~Deka, ``A deep framework for enhancement of
  diagnostic information in {CSMRI} reconstruction,'' \emph{Biomedical Signal
  Processing and Control}, vol.~71, p. 103117, Jan. 2022.

\bibitem{Wang2016b}
S.~Wang, Z.~Su, L.~Ying, X.~Peng, S.~Zhu, F.~Liang, D.~Feng, and D.~Liang,
  ``Accelerating magnetic resonance imaging via deep learning,'' in \emph{Proc.
  ISBI}, Apr. 2016, pp. 514--517.

\bibitem{Jin2017}
K.~H. Jin, M.~T. McCann, E.~Froustey, and M.~Unser, ``Deep convolutional neural
  network for inverse problems in imaging,'' \emph{IEEE Trans. Image Process.},
  vol.~26, no.~9, pp. 4509--4522, Jun. 2017.

\bibitem{Quan2018}
T.~M. Quan, T.~Nguyen-Duc, and W.-K. Jeong, ``Compressed sensing {MRI}
  reconstruction using a generative adversarial network with a cyclic loss,''
  \emph{IEEE Trans. Med. Imag.}, vol.~37, no.~6, pp. 1488--1497, Mar. 2018.

\bibitem{Yang2018}
G.~Yang, S.~Yu, H.~Dong, G.~Slabaugh, P.~L. Dragotti, X.~Ye, F.~Liu,
  S.~Arridge, J.~Keegan, Y.~Guo, and D.~Firmin, ``{DAGAN}: Deep de-aliasing
  generative adversarial networks for fast compressed sensing {MRI}
  reconstruction,'' \emph{IEEE Trans. Med. Imag.}, vol.~37, no.~6, pp.
  1310--1321, Dec. 2018.

\bibitem{HNWei2022a}
H.~Wei, Z.~Li, S.~Wang, and R.~Li, ``Undersampled multi-contrast mri
  reconstruction based on double-domain generative adversarial network,''
  \emph{IEEE J. Biomed. Health Informat.}, pp. 1--1, 2022.

\bibitem{Oezbey2022}
M.~Özbey, O.~Dalmaz, S.~U. Dar, H.~A. Bedel, Şaban Özturk, A.~Güngör, and
  T.~Çukur, ``Unsupervised medical image translation with adversarial
  diffusion models,'' \emph{arXiv}, Oct. 2022.

\bibitem{Guengoer2022}
A.~Güngör, S.~U. Dar, Şaban Öztürk, Y.~Korkmaz, G.~Elmas, M.~Özbey, and
  T.~Çukur, ``Adaptive diffusion priors for accelerated mri reconstruction,''
  \emph{arXiv}, Nov. 2022.

\bibitem{Souza2019}
R.~Souza and R.~Frayne, ``A hybrid frequency-domain/image-domain deep network
  for magnetic resonance image reconstruction,'' in \emph{Proc. SIBGRAPI}, Oct.
  2019.

\bibitem{Zhou2020}
B.~Zhou and S.~K. Zhou, ``{DuDoRNet}: Learning a dual-domain recurrent network
  for fast {MRI} reconstruction with deep t1 prior,'' in \emph{Proc. IEEE Conf.
  Comput. Vis. Pattern Recognit. (CVPR)}, Jun. 2020, pp. 4272--4281.

\bibitem{Yang2020}
Y.~Yang, J.~Sun, H.~Li, and Z.~Xu, ``{ADMM}-{CSNet}: A deep learning approach
  for image compressive sensing,'' \emph{IEEE Trans. Pattern Anal. Mach.
  Intell.}, vol.~42, no.~3, pp. 521--538, Mar. 2020.

\bibitem{Wang2021}
S.~Wang, T.~Xiao, Q.~Liu, and H.~Zheng, ``Deep learning for fast {MR} imaging:
  A review for learning reconstruction from incomplete k-space data,''
  \emph{Biomedical Signal Processing and Control}, vol.~68, p. 102579, Jul.
  2021.

\bibitem{Aghabiglou2022}
A.~Aghabiglou and E.~M. Eksioglu, ``Deep unfolding architecture for {MRI}
  reconstruction enhanced by adaptive noise maps,'' \emph{Biomedical Signal
  Processing and Control}, vol.~78, p. 104016, Sep. 2022.

\bibitem{Gregor2010}
K.~Gregor and Y.~LeCun, ``Learning fast approximations of sparse coding,'' in
  \emph{Proc. ICML}, 2010, pp. 399--406.

\bibitem{Yang2016}
Y.~Yang, J.~Sun, H.~Li, and Z.~Xu, ``Deep admm-net for compressive sensing
  mri,'' in \emph{Proc. NeurIPS}, 2016.

\bibitem{Zheng2019}
H.~Zheng, F.~Fang, and G.~Zhang, ``Cascaded dilated dense network with two-step
  data consistency for mri reconstruction,'' in \emph{Proc. NeurIPS}, 2019.

\bibitem{Dar2023}
S.~U. Dar, Şaban Öztürk, M.~Özbey, and T.~Çukur, ``Learning deep mri
  reconstruction models from scratch in low-data regimes,'' \emph{arXiv}, Jan.
  2023.

\bibitem{Aggarwal2019}
H.~K. Aggarwal, M.~P. Mani, and M.~Jacob, ``{MoDL}: Model-based deep learning
  architecture for inverse problems,'' \emph{IEEE Trans. Med. Imag}, vol.~38,
  no.~2, pp. 394--405, Aug. 2019.

\bibitem{Duan2019}
J.~Duan, J.~Schlemper, C.~Qin, C.~Ouyang, W.~Bai, C.~Biffi, G.~Bello,
  B.~Statton, D.~P. O'Regan, and D.~Rueckert, ``{VS}-net: Variable splitting
  network for accelerated parallel {MRI} reconstruction,'' in \emph{Proc.
  MICCAI,}, Oct. 2019, pp. 713--722.

\bibitem{Schlemper2018}
J.~Schlemper, J.~Caballero, J.~V. Hajnal, A.~N. Price, and D.~Rueckert, ``A
  deep cascade of convolutional neural networks for dynamic {MR} image
  reconstruction,'' \emph{IEEE Trans. Med. Imag.}, vol.~37, no.~2, pp.
  491--503, Oct. 2018.

\bibitem{Zhang2018}
J.~Zhang and B.~Ghanem, ``{ISTA}-net: Interpretable optimization-inspired deep
  network for image compressive sensing,'' in \emph{Proc. IEEE Conf. Comput.
  Vis. Pattern Recognit. (CVPR)}, Jun. 2018, pp. 1828--1837.

\bibitem{Lustig2007}
M.~Lustig, D.~Donoho, and J.~M. Pauly, ``Sparse {MRI}: The application of
  compressed sensing for rapid {MR} imaging,'' \emph{Magn Reson Med.}, vol.~58,
  no.~6, pp. 1182--1195, Dec. 2007.

\bibitem{Xiang2020}
J.~Xiang, Y.~Dong, and Y.~Yang, ``{FISTA}-net: Learning a fast iterative
  shrinkage thresholding network for inverse problems in imaging,'' \emph{IEEE
  Trans. Med. Imag.}, vol.~40, no.~5, pp. 1329--1339, May 2021.

\bibitem{He2016}
K.~He, X.~Zhang, S.~Ren, and J.~Sun, ``Deep residual learning for image
  recognition,'' in \emph{Proc. IEEE Conf. Comput. Vis. Pattern Recognit.
  (CVPR)}, jun. 2016, pp. 770--778.

\bibitem{He2016a}
------, ``Identity mappings in deep residual networks,'' in \emph{Proc. ECCV},
  Sep. 2016, pp. 630--645.

\bibitem{Zhang2017}
X.~Zhang, Z.~Li, C.~C. Loy, and D.~Lin, ``{PolyNet}: A pursuit of structural
  diversity in very deep networks,'' in \emph{Proc. IEEE Conf. Comput. Vis.
  Pattern Recognit. (CVPR)}, Jul. 2017, pp. 3900--3908.

\bibitem{Gomez2017}
A.~Gomez, M.~Ren, R.~Urtasun, and R.~Grosse, ``The reversible residual network:
  Backpropagation without storing activations,'' in \emph{Proc. NeurIPS}, 2017.

\bibitem{Lu2018}
Y.~Lu, A.~Zhong, Q.~Li, and B.~Dong, ``Beyond finite layer neural networks:
  Bridging deep architectures and numerical differential equations,'' in
  \emph{Proc. ICML}, vol.~80, Aug. 2018, pp. 3276--3285.

\bibitem{Katrutsa2017}
A.~Katrutsa, T.~Daulbaev, and I.~Oseledets, ``Deep multigrid: learning
  prolongation and restriction matrices,'' \emph{arXiv}, Nov. 2017.

\bibitem{Ke2017}
T.-W. Ke, M.~Maire, and S.~X. Yu, ``Multigrid neural architectures,'' in
  \emph{Proc. IEEE Conf. Comput. Vis. Pattern Recognit. (CVPR)}, Jul. 2017, pp.
  4067--4075.

\bibitem{He2019}
J.~He and J.~Xu, ``{MgNet}: A unified framework of multigrid and convolutional
  neural network,'' \emph{Sci. China Math.}, vol.~62, no.~7, pp. 1331--1354,
  May 2019.

\bibitem{Fan2021}
X.~Fan, Y.~Yang, and J.~Zhang, ``Deep geometric distillation network for
  compressive sensing {MRI},'' in \emph{Proc. IEEE {EMBS} Int. Conf. BHI}, Jul.
  2021.

\bibitem{McCormick1987}
S.~F. McCormick, ``Multigrid methods,'' \emph{Society for Industrial and
  Applied Mathematics}, 1987.

\bibitem{KChen05}
K.~Chen, \emph{Matrix Preconditioning Techniques and Applications}.\hskip 1em
  plus 0.5em minus 0.4em\relax Cambridge University Press, 2005.

\bibitem{You2021}
D.~You, J.~Xie, and J.~Zhang, ``Ista-net++: Flexible deep unfolding network for
  compressive sensing,'' in \emph{Proc. IEEE Int. Conf. ICME}, Jul. 2021.

\bibitem{Wang2018}
X.~Wang, K.~Yu, S.~Wu, J.~Gu, Y.~Liu, C.~Dong, Y.~Qiao, and C.~C. Loy,
  ``{ESRGAN}: Enhanced super-resolution generative adversarial networks,'' in
  \emph{Proc. ECCV}, Jan. 2019, pp. 63--79.

\bibitem{Zhang2018a}
Y.~Zhang, Y.~Tian, Y.~Kong, B.~Zhong, and Y.~Fu, ``Residual dense network for
  image super-resolution,'' in \emph{Proc. IEEE Conf. Comput. Vis. Pattern
  Recognit. (CVPR)}, Jun. 2018, pp. 2472--2481.

\bibitem{Zhang2020}
K.~Zhang, L.~V. Gool, and R.~Timofte, ``Deep unfolding network for image
  super-resolution,'' in \emph{Proc. IEEE Conf. Comput. Vis. Pattern Recognit.
  (CVPR)}, Jun. 2020, pp. 3214--3223.

\bibitem{Glorot2011}
X.~Glorot, A.~Bordes, and Y.~Bengio, ``Deep sparse rectifier neural networks,''
  in \emph{Proc. Int. Conf. Artif. Intell. and Statist.}, vol.~15, Jan. 2011,
  pp. 315--323.

\bibitem{KJiang2018}
K.~Jiang, Z.~Wang, P.~Yi, J.~Jiang, J.~Xiao, and Y.~Yao, ``Deep distillation
  recursive network for remote sensing imagery super-resolution,'' \emph{Remote
  Sens.}, vol.~10, no.~11, p. 1700, Oct. 2018.

\bibitem{Lee2015}
C.-Y. Lee, S.~Xie, P.~Gallagher, Z.~Zhang, and Z.~Tu, ``{Deeply-Supervised
  Nets},'' in \emph{Proc. PMLR}, vol.~38, Aug. 2015, pp. 562--570.

\bibitem{He2015}
K.~He, X.~Zhang, S.~Ren, and J.~Sun, ``Delving deep into rectifiers: Surpassing
  human-level performance on {ImageNet} classification,'' in \emph{Proc. ICCV},
  Dec. 2015, pp. 1026--1034.

\bibitem{NNA2015}
\BIBentryALTinterwordspacing
A.~Andrew, A.~Akhondi-Asl, W.~Hongzhi, T.~Nicholas, A.~Brian, W.~S. K, and
  L.~Bennett, ``Miccai 2013 segmentation algorithms, theory and applications
  (sata) challenge results summary,'' in \emph{MICCAI Challenge Workshop on
  Segmentation: Algorithms, Theory and Applications (SATA)}, 2013. [Online].
  Available: \url{https://masi.vuse.vanderbilt.edu/submission/leaderboard.html}
\BIBentrySTDinterwordspacing

\bibitem{Marcus2007}
D.~S. Marcus, T.~H. Wang, J.~Parker, J.~G. Csernansky, J.~C. Morris, and R.~L.
  Buckner, ``Open access series of imaging studies ({OASIS}): Cross-sectional
  {MRI} data in young, middle aged, nondemented, and demented older adults,''
  \emph{Journal of Cognitive Neuroscience}, vol.~19, no.~9, pp. 1498--1507,
  Sep. 2007.

\bibitem{Andreopoulos2008}
A.~Andreopoulos and J.~K. Tsotsos, ``Efficient and generalizable statistical
  models of shape and appearance for analysis of cardiac {MRI},'' \emph{Med.
  Image Anal.}, vol.~12, no.~3, pp. 335--357, Jun. 2008.

\bibitem{Kingma2014}
D.~Kingma and J.~Ba, ``Adam: A method for stochastic optimization,'' in
  \emph{Proc. ILCR}, Dec. 2014.

\bibitem{Bernstein2001}
M.~A. Bernstein, S.~B. Fain, and S.~J. Riederer, ``Effect of windowing and
  zero-filled reconstruction of {MRI} data on spatial resolution and
  acquisition strategy,'' \emph{J. Magn. Reson. Imaging}, vol.~14, no.~3, pp.
  270--280, Sep. 2001.

\end{thebibliography}

\end{document}